\documentclass[twocolumn,amsmath,amssymb,aps, prb,
showkeys,showpacs, superscriptaddress] {revtex4-1}
\usepackage{graphicx}
\usepackage{bm}
\usepackage[colorlinks=true,breaklinks=true,allcolors=blue]{hyperref}
\usepackage{dcolumn}
\usepackage{slashed}

\allowdisplaybreaks

\begin{document}
\title{Quantum Electronic Circuit Simulation of Generalized sine-Gordon Models}
\author{Ananda Roy}
\email{ananda.roy@tum.de}
\affiliation{Institut de Physique Th\'eorique, Paris Saclay University, CEA, CNRS, F-91191 Gif-sur-Yvette, France}
\affiliation{Department of Physics, T42, Technische Universit\"at M\"unchen, 85748 Garching, Germany}
\author{Hubert Saleur}
\affiliation{Institut de Physique Th\'eorique, Paris Saclay University, CEA, CNRS, F-91191 Gif-sur-Yvette, France}
\affiliation{Department of Physics and Astronomy, University of Southern California, Los Angeles, CA, USA}
\begin{abstract}
Investigation of strongly interacting, nonlinear quantum field theories (QFT-s) remains one of the outstanding challenges of modern physics. Here, we describe analog quantum simulators for nonlinear QFT-s using mesoscopic superconducting circuit lattices. Using the Josephson effect as the source of nonlinear interaction, we investigate generalizations of the quantum sine-Gordon model. In particular, we consider a two-field generalization, the double sine-Gordon model. In contrast to the sine-Gordon model, this model can be {\it purely quantum integrable}, when it does not admit a semi-classical description - a property that is generic to many multi-field QFT-s. The primary goal of this work is to investigate different thermodynamic properties of the double sine-Gordon model and propose experiments that can capture its subtle quantum integrability. First, we analytically compute the mass-spectrum and the ground state energy in the presence of an external `magnetic' field using Bethe ansatz and conformal perturbation theory. Second, we calculate the thermodynamic Bethe ansatz equations for the model and analyze its finite temperature properties. Third, we propose experiments to verify the theoretical predictions. 
\end{abstract}
\maketitle 

\section{Introduction}
\label{intro}
The longstanding goal of quantum field theory (QFT) is to predict the masses of the excitations and their scattering cross-sections in terms of the parameters of the theory and to characterize the different phases and the phase-transition points. While remarkable progress has been achieved in analyzing strongly coupled QFT-s using numerical or effective field theory methods, many quantities of interest remain elusive, either due to the computational limitations or due to the lack of an effective field theory. Quantum simulation, both analog and digital, takes a different approach to solving these aforementioned problems, where one quantum system is tailored to simulate another in a controlled manner~\cite{Feynman_1982,Lloyd_1996,Lloyd1997, Doucot2004,Buchler2005,Cirac2010,Casanova2011,Jordan2011,Jordan2012}. 

Here, we describe analog quantum simulators for some QFT-s with superconducting quantum electronic circuit (QEC) lattices~\cite{Houck2012,Devoret_Schoelkopf_2013}. Specifically, we are interested in obtaining  bosonic QFT-s directly, and  not  as consequences of mathematical manipulations  (bosonization) of an underlying fermionic or spin system~\cite{Fendley1995, Fendley1995a, Milliken1996, Lesage1997, Giulliano2005, Gritsev2007}. Even in $1+1$ space-time dimensions, this is crucial to distill the true behavior of a bosonic system from a bosonized fermionic system~\cite{Bukhvostov1980, Saleur1999}, despite the well-known fermion-boson correspondences~\cite{Giamarchi2003, Gogolin2004}. Of course, this distinction is even more important in higher dimensions. 

Attempts in this direction have been considered previously in the well-established Bose-Hubbard model paradigm~\cite{Fisher1989, Sachdev2011}. Another scheme is to use the Josephson effect  to give rise naturally to the cosine potential~\cite{Tinkham2004} of the  sine-Gordon (SG) model (see more below). Our construction generalizes the latter strategy.  The role of the bosonic field at a point in space-time is played by the time-integral of the voltage at that point~\cite{Devoret_1997} (so the  underlying degrees of freedom of the QEC lattices are faithful directly to the bosonic description). The building blocks of the QEC lattices are superconducting self and mutual inductances, capacitances and Josephson junctions. The current-voltage constitutive relations of these elements, together with Kirchhoff's laws for the circuits, give rise to the nonlinear field equations of the QFT-s. 

The playground for  bosonic nonlinear QFT-s is wide, even in $1+1$ dimensions, and includes typically cousins of the SG theory involving multiple component fields, together with some cosine interaction. The case of two bosons already escapes our complete understanding. In contrast with the SG case, the physics of the various ``double sine-Gordon models" (see definitions below) cannot always be inferred from some classical limit. For instance, fixed points are known to exist, which are truly quantum in nature and do not admit a semi-classical description \footnote{This is true also if one tries to ``backtrack'' to the fermionic formulation,  where ``non-Fermi liquid'' fixed points are then encountered}. In fact, one of the simplest new aspects to formulate and study in multi-component bosonic theories is integrability. Integrability leads to a factorized multi-particle scattering matrix~\cite{Iagolnitzer1978,Zamolodchikov1979}. This, in turn, allows analytic computation of thermodynamic and transport properties of these QFT-s~\cite{Zamolodchikov1990a,Smirnov1992}. The ability to do such computations is crucial to understand non-perturbative aspects of strongly interacting systems - for instance,  charge fractionalization~\cite{Senthil2000, Xu2010, Roy2017} and its effect on full-counting statistics~\cite{Nazarov_Blanter_2009}, both topics of high current interest.

It is well known that the bulk and boundary SG models are  classically integrable~\cite{Rajaraman1982,Ablowitz2006}. This was largely used historically to establish their  quantum  integrablity as well ~\cite{Mandelstam1975, Zamolodchikov1979,Ghoshal1994}. While  the  naive intuition that quantum fluctuations might well  destroy classical integrability - as is known to be the case, for instance, for many sigma models~\cite{Zarembo2017}- is usually correct,  it is now understood that  the SG model remains integrable in the quantum regime due to the existence of subtle quantum group symmetries~\cite{Bernard1991}. The presence of these symmetries  in $1+1$ dimensions relies heavily on the existence of conformally invariant UV fixed points, of which integrable QFTs can be considered as deformations~\cite{Belavin1984, Zamolodchikov1987,Zamolodchikov1989, diFrancesco1997,Mussardo2010}.   

The simplest generalization of the SG model in this regard is probably the double sine-Gordon  (dSG) model, whose  euclidean action is   
\begin{align}
\label{dSGaction}
{\cal S}_{\rm{dSG}} =\int d^2x \frac{1}{2}\sum_{i=1,2}(\partial_\mu\phi_i)^2 + \frac{2M_0}{\pi}\prod_{i=1,2}\cos(\alpha_i \phi_i),
\end{align}
where $\phi_1, \phi_2$ are two bosonic fields, $M_0$  the interaction strength and $\alpha_1, \alpha_2$ the coupling constants. This model and its close cousins involving a boundary interaction appear in a variety of guises in condensed matter physics - for instance, in transport experiments involving one-dimensional particles with charge and spin, or in quantum Brownian motion on two dimensional lattices~\cite{Affleck2001}. Unfortunately, little remains known about the possibility to solve the dSG model exactly, and to infer from such a solution properties of physical interest. This is because, remarkably, as soon as more than one bosonic degree of freedom is involved, quantum and classical integrability often part ways. In fact, it is  known that the dSG model is classically integrable only for $\alpha_1 = \alpha_2$, when it reduces to two decoupled SG models~\cite{Ameduri1998}. However, the same fluctuations that can destroy classical integrability in some cases can also  give rise to quantum integrability.   Purely quantum integrable manifolds, if they exist, should thus arise for values of the coupling constants  that cannot approach the origin in the  $(\alpha_{1}, \alpha_{2})$ plane.  The dSG model is probably only quantum integrable on several $(\alpha_1,\alpha_2)$ manifolds~\cite{Bukhvostov1980,Fateev1996,Lesage1997,Lesage1998,Ameduri1998a, Bazhanov2018}. While some doubts remain on the exact nature of this statement, it is strongly believed in the community that  the manifold: $\alpha_1^2 + \alpha_2^2 = {4\pi/\hbar}$ indeed is quantum integrable (in the following we will set $\hbar=1$). This is, of course, a remarkable statement: we are facing a situation where  a {\it classical} soliton wave-packet, which solves the classical field equations, gets scrambled, but its {\it quantized} counterpart, when the couplings reach some magical values,  propagates undistorted and scatters only with phase-shifts! One of our goals is to propose a set-up where such behavior - which, we believe, is the norm rather than the exception in many  multi-field  theories - might be observed experimentally. 

It is important to stress that the calculations presented in this paper are done exactly for the model in Eq. \eqref{dSGaction}. It should be contrasted with those done by extrapolating the results of the more general Fateev model~\cite{Fateev1996}, of which the dSG is a special case. However, the Fateev model involves, on top of the two (compact) bosonic fields $\phi_i$, an extra non-compact bosonic field. The presence of this other field  makes many calculations of the Fateev model - in particular, the thermodynamic Bethe ansatz (TBA) - considerably easier. It is expected that the final results can then be continued to the limit where this extra boson decouples, but this assumption is dangerous in view of the singularity of this limit. 

\section{The double sine-Gordon model at zero temperature}
\label{zerotemp}
The solitons of the quantum dSG model cannot be approached with semiclassical methods, and are not fully understood. It is known that they are topological excitations of the fields $\varphi_1, \varphi_2$, where $\varphi_{1,2} = (\alpha_1\phi_1 \pm\alpha_2\phi_2)/2\sqrt{\pi}$. They carry a pair of quantum numbers corresponding to the fields $\phi_1, \phi_2$, just as an electron carries  electric charge and spin quantum numbers. These solitons scatter with a factorized scattering matrix. More remarkably, because of the underlying (pair of) quantum group symmetries~\cite{Lesage1998}, the two quantum numbers of each soliton scatter independently of each other. The factorized scattering matrix is  expected to be given by~\cite{Fateev1996} $S = S_{p_1}\otimes S_{p_2}$, where $S_{p_i}$, $i=1,2$ is the SG scattering matrix. The corresponding SG couplings $\beta_i$-s are related to $p_i$-s by $\beta_i^2/8\pi = p_i/(p_i+1)$, $i = 1,2$~\cite{Mussardo2010,Zamolodchikov1979}, where $p_{1,2} = \alpha_{1,2}^2/2\pi$~\cite{Fateev1996}. 

The masses of the physical excitations in terms of the parameters of the action, together with  a verification of the scattering matrix, are obtained by calculating the ground state energy in the presence of `magnetic fields' $h_{1,2}$ coupling to the conserved charges $Q_{1,2} = \int dx \partial_x\varphi_{1,2}/\sqrt{\pi}$. Consider the limit $h_{1,2}\gg M_0$ and tune $h_2$ such that $\langle Q_2\rangle=0$. Standard Bethe ansatz calculations using Wiener-Hopf technique~\cite{Fateev1996, Fateev1993} show that, on the integrable manifold, the ground state energy gets a logarithmic correction in $h_1$ (to be contrasted with the purely polynomial corrections of the Fateev model~\cite{Fateev1996}), in addition to its usual quadratic dependence: 
\begin{equation}
{\cal E}(h_1) = -\frac{h_1^2p_1(2-p_1)}{2\pi} +\frac{m_s^2}{2\pi}\sin^2\Big(\frac{\pi p_1}{2}\Big)\ln h_1.
\end{equation}
Here $m_s$ is the mass of the $\varphi_1$ soliton~\cite{sol_foot}. It is obtained in terms of the action parameters using conformal perturbation theory (CPT)~\cite{Klassen1991}. We find that $m_s = M_0/\sin(\pi p_1/2)$, which matches the result of Ref. \onlinecite{Fateev1996}. The calculation is is described in detail in Appendix \ref{Wiener-Hopf}. 

\section{The double sine-Gordon model at nonzero temperature}
\label{tba}
The thermodynamics of QFT-s in an infinite volume can be inferred from its scattering matrix~\cite{Dashen1969} using the TBA technique~\cite{Zamolodchikov1990a}. Using the factorized scattering matrix given above, we now compute the free energy of the dSG model at a finite temperature $1/R$. The TBA given below is essential to correctly predict the dSG free energy and cannot be directly  inferred from the TBA of the Fateev model (the free energy extrapolated  from the Fateev model will appear to be infinite). Without loss of generality, we choose $p_1< 1$, $i.e.,$ the first SG sector to be attractive, and $p_2 > 1$, $i.e.,$ the second, repulsive. In addition to the solitons, the spectrum of the theory also includes $n < 1/p_1$ particles which are neutral with respect to first quantum number, but still carry the second quantum number. They are the bound states of the fundamental `double' solitons and correspond to the ordinary breather poles of the attractive SG scattering amplitude of the first sector~\cite{Zamolodchikov1979}. The TBA analysis is, in general, highly non-trivial, the scattering in each sector being non-diagonal. We consider the simpler case $p_1 = 1/(n + 1)$ when the first sector scattering is diagonal. The diagonalization problem of the second sector is done by the Takahashi-Suzuki classification of the solution of the Bethe equations~\cite{Takahashi1972,Fendley1992,Takahashi2005} and the algebraic Bethe ansatz~\cite{Korepin1993,Faddeev1996}.  The detailed derivation of the TBA equations is given in Appendix \ref{TBA}. The resulting TBA equations have an universal form for all $n$: 
\begin{align}
\label{TBA_eq}
\epsilon_\gamma(\theta) &= \hat\varphi_n \star {\cal L}_{b_n},\ \gamma = s,a, \nonumber\\
\epsilon_{b_n}(\theta) &= \hat\varphi_n \star ({\cal L}_s + {\cal L}_a + {\cal L}_{b_{n-1}}),\nonumber \\
\epsilon_{b_j}(\theta) &= \hat\varphi_n \star({\cal L}_{b_{j+1}} + {\cal L}_{b_{j-1}}), \ j = n-1, \ldots, 3,\nonumber\\
\epsilon_{b_2}(\theta) &= \hat\varphi_n \star({\cal L}_{b_{3}} + {\cal L}_{b_{1}} - {\cal L}_1), \nonumber\\	
\epsilon_{b_1}(\theta) &= \hat\varphi_n \star({\cal L}_{b_{2}}  - {\cal L}_2),\nonumber \\
\epsilon_{1}(\theta) &= -m_{b_1}R\cosh\theta + \hat\varphi_n \star({\cal L}_{b_{2}}  - {\cal L}_2),\nonumber\\
\epsilon_{2}(\theta) &= -\hat\varphi_n \star(L_{b_{1}}  - L_1 + L_3),\nonumber\\
\epsilon_{j}(\theta) &= -\hat\varphi_n \star (L_{j+1} + L_{j-1}), \ j = n-1, \ldots, 3,\nonumber\\
\epsilon_n(\theta) &= -\hat\varphi_n \star(L_{n-1} + L_{n+1} + L_{n+2}), \nonumber\\
\epsilon_{k}(\theta) &= -\hat\varphi_n \star L_{n},\ k = n+1, n+2.
\end{align}
The TBA kernel is $\hat\varphi_n(\theta) = (n+1)/\cosh[(n+1)\theta]$, while  ${\cal L}(\theta) = \ln[1 + e^{\epsilon(\theta)}]$ and $L(\theta) = \ln[1 + e^{-\epsilon(\theta)}]$. Finally, $m_{b_1} = 2m_s\sin{\pi p_1/2}$ is the mass of the first breather~\cite{Zamolodchikov1979} and $a \star b = \int a(\theta-\theta') b(\theta')d\theta'/2\pi$. The TBA diagram is given in Fig. \ref{TBA_pic}. 

\begin{figure}
\centering
\includegraphics[width = 0.475\textwidth]{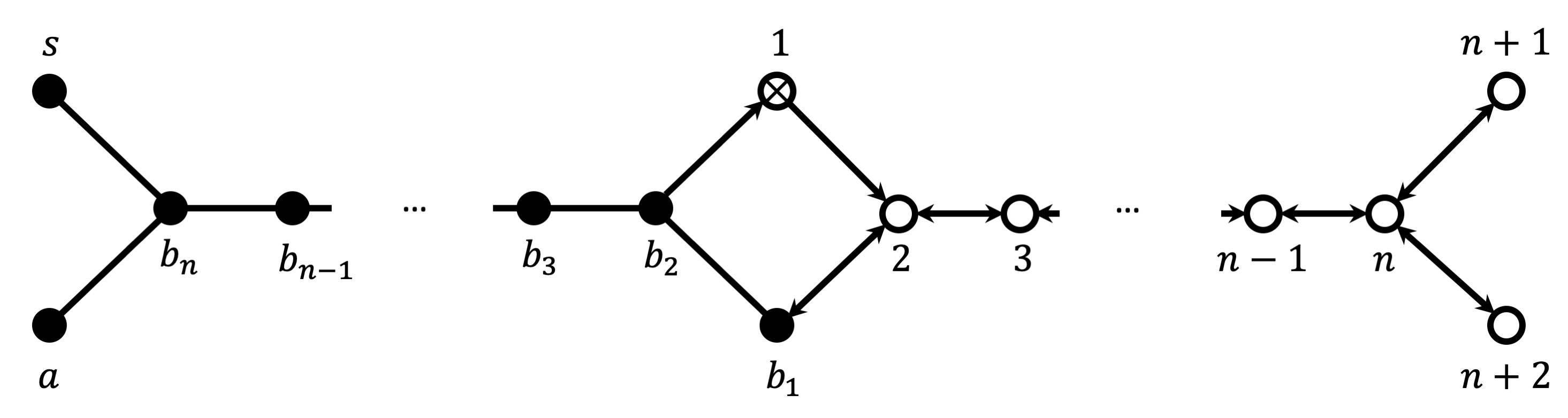}
\caption{\label{TBA_pic}TBA diagram for the dSG model. The solid circles denote the physical massive particles: soliton ($s$), antisoliton($a$) and breathers ($b_j$, $j = 1,\ldots, n$). The empty circles denote pseudoparticles ($j=1,\ldots, n+2$) needed to diagonalize the dSG scattering matrix. The pseudoparticle $1$ has a cross on it, which indicates that this particle has a mass term in its TBA equation. The connectivity of the diagram encodes which particles show up in the  TBA equation of a given particle. The arrows on the links encode the sign of the term on the right hand side (RHS) of the TBA equation. For instance, for a link connecting particles $p, q$, if there is an arrow incident on $q$ and none on $p$, then the RHS of the TBA equation for $p$ has a minus sign in front of the term involving $q$, while the term on the RHS for $q$ involving $p$ has a plus sign. }
\end{figure}

We check our TBA equations by analytic computation of the central charge in the conformal (UV) limit. This yields the expected result of $c_{\rm{UV}} = 2$ for the two bosonic fields (see Appendix \ref{TBA} for details of the computation). Next, we compute the free energy of the system upon deviation from the conformal limit. This is given in terms of the effective central charge, $c_{\rm{eff}} = -6R^2f/\pi$, where $f$ is the free energy per unit length. On the integrable manifold, the perturbation has dimension  $(\alpha_1^2 + \alpha_2^2)/(4\pi)=1$. This leads to logarithmic corrections to $c_{\rm{eff}}$~\cite{Zamolodchikov1991a, Fendley1993} upon deviations from the conformal limit, in addition to the polynomial corrections in powers of $(m_sR)^2$~\cite{Klassen1991}. We calculate these corrections by numerically solving the TBA equations. The logarithmic correction for the different values of $n$ is then verified using CPT. We find the logarithmic correction to $c_{\rm{eff}}$ to be
\begin{equation}
(\Delta c_{\rm{eff}})_{\rm{log}} = \frac{6(m_sR)^2}{\pi^2}\sin^2\Big[\frac{\pi}{2(n+1)}\Big]\ln(m_sR).
\end{equation}
More details on the TBA and the explicit forms of the different polynomial corrections to $c_{\rm{eff}}$ for $n=2,3$ are given in Appendix \ref{TBA}. Note that this logarithmic correction is very similar to those obtained for the sine-Gordon model for suitably chosen coupling constants ~\cite{Zamolodchikov1991a, Fendley1993} and should be contrasted with a very different type of logarithmic correction that arises in the Fateev or sinh-Gordon models~\cite{Fateev1996, Zamolodchikov2006}. 

\section{The double sine-Gordon model with quantum circuits}
\begin{figure}
\centering
\includegraphics[width = 0.48\textwidth]{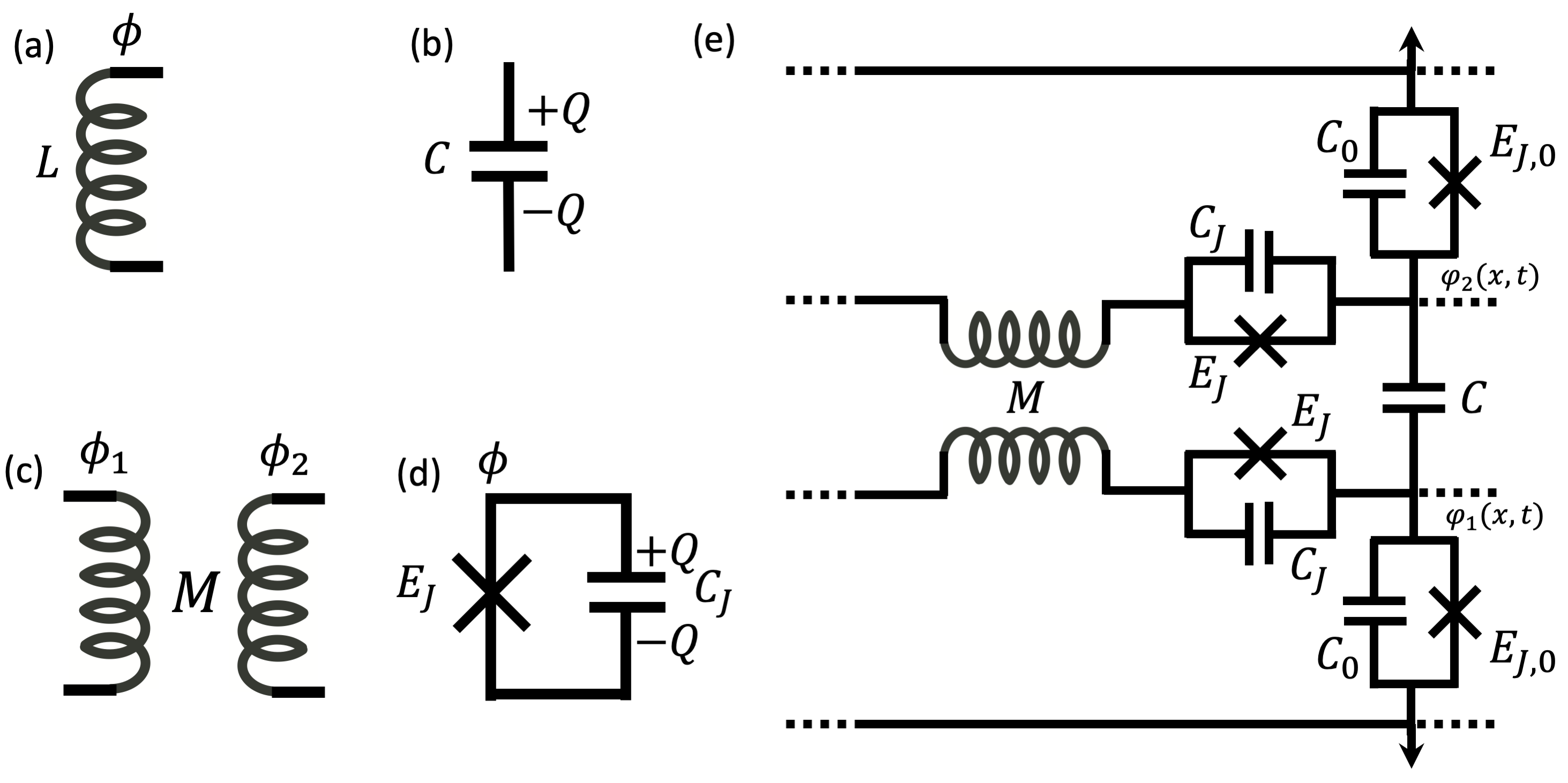}
\caption{\label{dsg_pic} (a) - (d): Essential primitives of QEC lattices [(a) an inductance, $L$, (b) a capacitance, $C$, (c) a mutual inductance, $M$, and (d) a Josephson junction (junction energy $E_J$ and junction capacitance $C_J$)]. The magnetic fluxes going through the inductance ($L$), the mutual inductance ($M$) and the Josephson junction are denoted by $\phi$, $\phi_1, \phi_2$ and $\phi$ respectively.  The charges on the  capacitor plates of the capacitances $C$ and $C_J$ are denoted by $Q$. (e) One unit cell of the QEC for the dSG model. The top and bottom horizontal lines denote the electrical ground. The Josephson junctions (junction energy $E_{J,0}$ and junction capacitance $C_0$) on the vertical links give rise to the nonlinear interaction. The Josephson junctions (junction energy $E_J$ and junction capacitance $C_J$) give rise to a large impedance of the array, without any nonlinear effects. Finally, the mutual inductance $M$ and the capacitance $C$ provide the necessary inductive and capacitive coupling between the upper and lower parts of the array. The bosonic fields, $\varphi_1,\varphi_2$, are the node-fluxes at the points indicated. }
\end{figure}
The predictions for the various thermodynamic quantities computed above can, in principle, be measured in an experimental setup. In particular, experiments should be able to capture the subtle quantum vs classical integrability of the dSG model. To that end, we provide a QEC for the dSG model. In terms of the rotated fields $\varphi_{1,2}$, the dSG action can be viewed as two SG actions coupled by $B\int d^2x \partial_\mu\varphi_1\partial_\mu\varphi_2$, where $B = \pi/\alpha_1^2 -\pi/\alpha_2^2$. Recall that in QEC-s, the bosonic field is time-integral of the voltage at a point. Thus, the first coupling term, $\partial_t\varphi_1\partial_t\varphi_2$, is a voltage-voltage coupling, realized in QEC-s with a capacitance ($C$). The second coupling term,  $\partial_x\varphi_1\partial_x\varphi_2$, is a current-current coupling, realized by a mutual inductance ($M$). All that is left is to realize two separate identical SG models. The QEC-s for the two SG models are two one-dimensional arrays [see panel (e) of Fig. \ref{dsg_pic}] with Josephson junctions on horizontal links (junction energy $E_J$ and junction capacitance $C_J$)  together with Josephson junctions on vertical links (junction energy $E_{J,0}$ and junction capacitance $C_0$). We work in the regime when $E_J\gg E_{C_J}$ and $E_{C_0}\gg E_{C_J}$, where $E_{C_J} = (2e)^2/2C_J$ and $E_{C_0} = (2e)^2/2C_0$. The array impedance is $Z = R_Q\sqrt{2E_{C_0}/E_{J}}/2\pi$~\cite{Manucharyan2009, Koch2009, LeHur2012, Goldstein2013}, where $R_Q(\simeq6.5\ k\Omega$) is the resistance quantum (in units of Cooper pairs). Only the kinetic inductance of the Josephson junctions on the horizontal links are used to increase the impedance of the array, with the phase-slip amplitudes are exponentially small $\sim e^{-\sqrt{E_J/E_{C_J}}}$~\cite{Koch_Schoelkopf_2007}. The Luttinger parameter is $K = 2Z/R_Q$. Capacitively and inductively coupling these two SG models results in the QEC for the dSG model. Choosing $M=C$, the correspondences between the circuit components and the dSG action are: $\alpha_1^2 = \pi K/2$, $\alpha_2^2 = \pi/(2/K + 4\pi C)$ and $M_0 = \pi E_{J,0}$. A more detailed analysis of the circuit is given in Appendix \ref{qcirc}. 

Next, we describe various experimental measurements of the dSG model possible in a QEC setup. Estimation of $\alpha_1,\alpha_2$ can be done by biasing with two current sources (the `magnetic fields' $h_1,h_2$ considered earlier)  and measuring the transmission of an input field through the dSG lattice. This provides an estimate of the average current, $\langle\partial_x\varphi_{1,2}\rangle$, flowing through the horizontal links in the circuit. Comparison to CPT calculations, which give $\langle \partial_x\varphi_{1,2}\rangle = [(\alpha_1\pm\alpha_2)h_1 + (\alpha_1 \mp \alpha_2)h_2]/4\pi^{3/2}$ to leading order, yields estimates of $\alpha_1,\alpha_2$. Classical integrability can be captured by scattering measurements of a classical sine-Gordon soliton wavepacket~\cite{Ablowitz2006} for the either $\varphi_1$ or $\varphi_2$ propagating through the array. Due to the presence of the mutual inductance and capacitance coupling to the upper and lower part of the array (Fig. \ref{dsg_pic}), the wavepacket will distort. However, on the quantum integrable manifold, a quantum soliton would propagate undistorted. A signature of the factorized scattering of the quantum soliton, can be obtained by measuring the specific heat~\cite{Guarcello2018}, which is predicted from the free energy computed above (recall that quantum integrability and the factorized scattering of the quantum solitons is central to the entire TBA computation). This measurement also provides the mass of the quantum soliton. More details on the specific heat measurement proposal is given in Appendix \ref{qcirc}.
Additional transport signatures of quantum integrability can be obtained from the boundary dSG model, which is realized by terminating the lattice shown in panel (e) of Fig. \ref{dsg_pic} with two impurity junctions, one each for the upper and lower parts of the array. The quantum transport properties can be calculated along the lines of Refs. \onlinecite{Lesage1997, Lesage1998} starting with our TBA. However, inclusion of the boundary in the TBA involves additional technical complications. We intend to report on this in a future publication. 

\section{Summary}
To summarize, we have provided analytical predictions of the thermodynamic properties of the dSG model and a QEC for simulating it. Faithful analog simulation of QFT-s using QEC-s open new possibilities in investigation of their non-perturbative properties. Radio-frequency measurements possible with QEC-s allow a more detailed verification of the QFT-s' properties. At the same time, QEC simulation of integrable QFT-s may be viewed as a method to benchmark these analog simulators by comparing experiments to analytical theoretical predictions. This leads to a systematic investigation of more general QFT-s, by systematically including perturbations which break integrability. Using the robust controllability and customizability of QEC-s, this can be achieved as described below. A massless free-boson QFT can be included by taking the SG circuit after removing the Josephson junctions on the vertical links [upper or lower half of panel (e) of Fig. \ref{dsg_pic}]. A massive free-boson QFT is obtained from the latter by adding linear inductances to the ground at each vertical link. Interactions can be realized by either inductive coupling or capacitive coupling or through Josephson junctions. For instance, interacting QFT-s may be obtained by specific arrangements of Josephson junctions in tailored geometries~\cite{Bergeal_Devoret_2010, Roy_Devoret_2015, four_mode_foot}. Fermionic modes may potentially also be included using the topological Josephson effect~\cite{Fu2010}. Finally, the QEC-s proposed in this work can be generalized to  simulate higher dimensional QFT-s (integrable and non-integrable)~\cite{Zamolodchikov1981}.

\section*{Acknowledgments}
A.R. acknowledges fruitful discussions with Jonathan Belletete, Michel Devoret, Hermann Grabert, Michal Pawelkiewicz, and A. Douglas Stone. A.R. acknowledges the support of the Alexander von Humboldt foundation. H.S. was supported by the ERC Advanced Grant NuQFT. 
H.S. thanks F. Lesage and P. Simonetti for (early)  collaboration on  this project, and T. Albash for discussions.

\appendix
 
\section{The double sine-Gordon model at zero temperature}
\label{Wiener-Hopf}
In this section we consider the dSG model in the presence of external magnetic fields $h_i$-s, which couple to the conserved charges $Q_i$-s of the solitons. First, we calculate the ground state energy in the limit $h_i\rightarrow\infty$ and then consider deviations from this limit using CPT. Subsequently, the results are compared with Bethe ansatz calculations. The charges, $Q_i$-s, of the solitons are given by 
\begin{equation}
Q_i = \int_{-\infty}^\infty dx \frac{1}{\sqrt{\pi}}\partial_1\varphi_i,\ i = 1,2, 
\end{equation} 
where $\varphi_{1,2} = (\alpha_1\phi_1 \pm\alpha_2\phi_2)/2\sqrt{\pi}$. The coupling to the magnetic field is given by the following action
\begin{eqnarray}
{\cal S}_{\rm{mag}} &=& -\frac{h_1}{\sqrt{\pi}}\int\ d^2x\partial_1\varphi_1 -\frac{h_2}{\sqrt{\pi}}\int\ d^2x\partial_1\varphi_2.
\end{eqnarray}
In limit $h_1/M_0, h_2/M_0\rightarrow \infty$, the interaction term ${\cal S}_{\rm{int}}$ can be neglected. The ground state energy can be computed by completing squares in a Gaussian integral and is given by 
 \begin{equation}
 {\cal E} = -\frac{(h_1^2 + h_2^2)A-2h_1h_2B}{2\pi(A^2-B^2)}. 
 \end{equation}
 The equilibrium values of the charges $Q_{1,2} = \partial{\cal E}/\partial h_{1,2}$ are given by 
 \begin{equation}
 \label{charges}
 Q_1 = \frac{h_1A-h_2B}{\pi(A^2-B^2)},\ Q_2 = \frac{h_2A-h_1B}{\pi(A^2-B^2)}. 
 \end{equation}
We will be considering the situation when the fields $h_1, h_2$ are so chosen that $Q_2=0$ which implies, from the above equation, $h_2 = Bh_1/A$. Then, the charge $Q_1$ is given by 
\begin{equation}
\label{q1}
Q_1 = \frac{h_1}{\pi A} = \frac{h_1\alpha_1^2\alpha_2^2}{\pi^2(\alpha_1^2 + \alpha_2^2)}
\end{equation}
 and the ground state energy is 
 \begin{equation}
 \label{CPT_gsenergy}
 {\cal E}_c = -\frac{h_1^2\alpha_1^2\alpha_2^2}{2\pi^2(\alpha_1^2 + \alpha_2^2)}. 
 \end{equation}
 Next, we look at finite $h_1/M_0, h_2/M_0$ and consider fluctuations of the fields $\varphi_{1,2}$ around their asymptotic values given by Eq. [\ref{charges}]. This is where our results differ from those obtained in Ref. \onlinecite{Fateev1996}. As will be shown below, the corrections to the limit $h_1/M_0, h_2/M_0\rightarrow \infty$ occur at the second order and lead to logarithmic corrections, while the same for the Fateev model occur at fourth order and give rise to polynomial corrections. We define
 \begin{equation}
 \varphi_i(x) = \sqrt{\pi}Q_i x_1 +\tilde\varphi_i(x),\ i=1,2,
 \end{equation}
 where $\langle\tilde\varphi_i\rangle=0$. The original fields $\phi_{1,2}$ can be written has 
 \begin{eqnarray}
 \phi_{1,2} =\frac{\pi}{\alpha_1}(Q_1 \pm Q_2)x_1 + \tilde\phi_{1,2},
 \end{eqnarray}
 where we have defined $\tilde{\phi_{1,2}} = \sqrt{\pi}(\tilde\varphi_1 \pm \tilde\varphi_2)/\alpha_1$. 
 Rewriting the action in terms of $\tilde{\phi}_{1,2}$ and considering the case when $Q_2=0$, we have $ {\cal S}_0 + {\cal S}_{\rm{mag}}  = \tilde{\cal S}_0 + LR {\cal E}_c$, where
 \begin{align}
 \tilde{\cal S}_0&= \frac{1}{2}\int\ d^2x \big\{(\partial_\mu \tilde\phi_1)^2 + (\partial_\mu \tilde\phi_2)^2 \big\},\nonumber\\
 {\cal E}_c &=- \frac{h_1^2\alpha_1^2 \alpha_2^2}{2\pi^2(\alpha_1^2 + \alpha_2^2)}.
 \end{align}
Here, $L$ is the spatial extent of the 1D system and $R$ is the inverse temperature. In the following, we will keep the fluctuating fields $\tilde\phi_{1,2}$ and evaluate the corrections due to the interaction term ${\cal S}_{\rm{int}}$ perturbatively. We consider the case when $Q_2=0$. The interaction term can be written as 
 \begin{equation}
 {\cal S}_{\rm{int}} = \frac{2M_0}{\pi}\int d^2x\ \cos(\pi Q_1x_1+ \alpha_1 \tilde\phi_1)\cos(\pi Q_1x_1 + \alpha_2\tilde\phi_2).\nonumber
 \end{equation}
 This leads to 
\begin{eqnarray}
\label{logZexp}
\ln Z &=& -{\cal E}_c LR + \ln Z_0 + \frac{1}{2}\langle {\cal S}_{\rm{int}}^2\rangle_0 
\end{eqnarray}
 where we have used the fact that $\langle {\cal S}_{\rm{int}}\rangle_0=0$, $Z_0 = \int {\cal D}\hat\phi_1{\cal D}\hat\phi_2 {\rm{exp}}(-\hat{\cal S}_0)$ and  the averages $\langle\ \rangle_0$ are with respect to $Z_0$. In order to calculate the correction term, we need the following formula for averages of the vertex operators~\cite{Zamolodchikov1995}
 \begin{widetext}
 \begin{eqnarray}
 \langle e^{i\alpha \phi(x_1)}\ldots e^{i\alpha\phi(x_n)}e^{-i\alpha \phi(y_1)}\ldots e^{-i\alpha\phi(y_n)}\rangle_0 &=& \frac{\prod_{i>j}^n(|x_i-x_j||y_1-y_j|)^\frac{\alpha^2}{2\pi}}{\prod_{i,j=1}^n|x_i-y_j|^\frac{\alpha^2}{2\pi}},
 \end{eqnarray}
 \end{widetext}
where the average is taken with respect to the free-field action. Performing the integral, on the integrable manifold: $\alpha_1^2 + \alpha_2^2 = 4\pi$, we get 
\begin{equation}
\label{eqcpt}
 \langle {\cal S}_{\rm{int}}^2\rangle_0 = \frac{M_0^2}{2\pi}LR\Big\{-2\gamma_E - 2\ln(2\pi |Q_1|)+ 2\ln\Big(\frac{R}{2a}\Big)\Big\},\nonumber
 \end{equation}

where $\gamma_E$ is the Euler constant and $a$ is a lattice cut-off. Using this result, the specific ground state energy, up to second order in $M_0$, is given by
\begin{eqnarray}
\label{e_cpt}
{\cal E} &=& -\frac{1}{L}\frac{\partial \ln Z}{\partial R}= {\cal E}_c +\frac{M_0^2}{2\pi}\ln(2\pi|Q_1|) + {\cal E}_0,
\end{eqnarray}
where 
\begin{equation}
\label{e_0}
{\cal E}_0 = -\frac{1}{L}\frac{\partial \ln Z_0}{\partial R} + \frac{M_0^2}{2\pi}\big\{\gamma_E + 1+\ln(R/2a)\big\}
\end{equation}
Next, we calculate the ground state energy using Bethe ansatz. Consider the case when the fields are chosen so that $Q_2=0$. Due to the presence of the magnetic field, the solitons of the $\varphi_1$ fields acquire an additional energy $-h_1Q_1$ and as long as $h_1$ is larger than the mass of the $\varphi_1$ solitons, these solitons fill up all possible states within a `Fermi interval' $-B< \theta< B$, where $\theta$ is the rapidity parameter and $B$ is the `Fermi wavevector'. The specific ground state energy is then given by~\cite{Polyakov1983, Wiegmann1984, Korepin1993}
\begin{equation}
{\cal E}_c = -\frac{m_s}{2\pi}\int_{-B}^{B}d\theta\cosh(\theta)\epsilon(\theta),
\end{equation}
where $m_s$ is the mass of the $\varphi_1$ soliton and the quasiparticle energies $\epsilon(\theta)$ satisfy 
\begin{equation}
\int_{-B}^B d\theta\tilde{K}_c(\theta-\theta') \epsilon(\theta')= h_1 - m_s\cosh(\theta),
\end{equation}
with the boundary condition $\epsilon(\pm B) = 0$. Here, 
\begin{equation}
\tilde K_c(\theta) = \delta(\theta) - \frac{1}{2\pi}\frac{d}{d\theta}[\delta_{p_1}(\theta) + \delta_{p_2}(\theta)],
\end{equation}
where $\delta_{p_i}$-s are soliton-soliton scattering phase-shifts (for instance, see Ref. \onlinecite{Fateev1996} for explicit forms). The Fourier transform of $\tilde{K}_c$ is given by  
\begin{equation}
K_c(\omega) = \frac{\sinh[\pi\omega (p_1 + p_2)/2]}{2\sinh(\pi\omega p_1/2)\sinh(\pi\omega p_2/2)}\tanh\Big(\frac{\pi\omega}{2}\Big). 
\end{equation}
The leading order energy contribution in the limit of $h_1/M_0, h_2/M_0\rightarrow\infty$ can be obtained by the standard manipulations~\cite{Fateev1996,Fateev1993} and the result is identical to that of the Fateev model. This leads to 
\begin{equation}\label{BA_gsenergy}
{\cal E}_c  = \frac{h_1^2}{2\pi K_c(0)} = -\frac{h_1^2p_1p_2}{\pi(p_1 + p_2)}. 
\end{equation}
Comparing Eqs. [\ref{CPT_gsenergy}, \ref{BA_gsenergy}], we get  
\begin{equation}
\label{pdef}
p_i = \frac{\alpha_i^2}{2\pi}, \ i = 1,2.
\end{equation}
In terms of $p_i$-s, the integration manifold is then given by $p_1 + p_2 = 2$. Next, we consider deviations from the $h_1/M_0, h_2/M_0\rightarrow \infty$ limit. This is done by a Wiener-Hopf calculation. Since we are interested in the dSG model on the integration manifold, we set $p_2 = 2-p_1$ at the outset. Then, the scattering kernel reduces to 
\begin{equation}
K_c(\omega) = \frac{\sinh(\pi\omega)}{2\sinh(\frac{\pi\omega p_1}{2})\sinh(\pi\omega - \frac{\pi\omega p_1}{2})}\tanh\Big(\frac{\pi\omega}{2}\Big).
\end{equation}
The kernel factorizes as $K_c(\omega) = 1/N(\omega)N(-\omega)$~\cite{Fateev1996}, where 
\begin{eqnarray}
N(\omega) = \sqrt{\frac{4\pi}{p_1(2-p_1)}}\frac{p(i\omega)p(\frac{i\omega}{2})e^{i\omega\Lambda}}{p(\frac{i\omega p_1}{2})p(i\omega - \frac{i\omega p_1}{2})p(\frac{1}{2} + \frac{i\omega}{2})},\nonumber
\end{eqnarray}
where \begin{equation}
\Lambda = -\ln 2 + \frac{2-p_1}{2}\ln(2-p_1) + \frac{p_1}{2}\ln p_1. 
\end{equation}
Then, the Bethe ansatz integral equation is reduced to the following linear integral equation for the $v(\omega)$, given by 
\begin{eqnarray}
\label{v_eq}
v(k) &=& -\frac{ih_1N(0)}{k} + \frac{i m_se^B}{2}\frac{N(-i)}{k-i}\nonumber\\&&+ \int_{C_+}\frac{d\omega}{2\pi i}\frac{e^{2i\omega B}}{w+k}\alpha(\omega)v(\omega),
\end{eqnarray}
where $m_s$ is the mass of physical excitations of solitons of $\varphi_1$, $\alpha(\omega) = N(\omega)/N(-\omega)$ and 
\begin{eqnarray}
\label{b_eq}
i h_1 N(0)&=& - \frac{i m_se^B}{2}N(-i)  \nonumber\\&& + \int_{C_+}\frac{d\omega}{2\pi i}e^{2i\omega B}\alpha(\omega)v(\omega).
\end{eqnarray}
Here, $C_+$ includes all the poles on the positive imaginary axis, but not the one at $\omega = -k$. Then, the ground state energy is given by 
\begin{eqnarray}
{\cal E}(h_1)-{\cal E}(0) &=& -\frac{h_1^2N(0)^2}{2\pi}\nonumber\\&& - \frac{h_1 N(0)}{\pi} \int_{C'_+}\frac{d\omega}{2\pi i}\frac{e^{2i\omega B}}{\omega - i}\alpha(\omega)v(\omega),\nonumber
\end{eqnarray}
where $C'_+$ is the contour that includes the poles included by $C_+$ plus the one at $\omega = i$. Our goal is to calculate this ground state energy, which gets contribution from the poles of $\alpha(\omega)$, $v(\omega)$ and $\omega = i$. We include these contributions iteratively. Keeping to the lowest order contribution, we evaluate the contribution due to the pole at $\omega = \omega_1 = i$. This leads to 
\begin{eqnarray}
{\cal E}(h_1)-{\cal E}(0) &=& -\frac{h_1^2N(0)^2}{2\pi} \nonumber\\&&+ \frac{h_1^2 N(0)^2}{\pi} \int_{C'_+}\frac{d\omega}{2\pi i}\frac{e^{2i\omega B}}{\omega(\omega - i)^3}\tilde\alpha(\omega),\nonumber
\end{eqnarray}
where $\tilde\alpha(\omega) = (\omega-i)\alpha(\omega)$. Application of the residue theorem leads to
\begin{eqnarray}
{\cal E}(h_1)-{\cal E}(0) &=& -\frac{h_1^2N(0)^2}{2\pi}\nonumber\\&& - \frac{h_1^2 N(0)^2}{2\pi}\frac{m_s^2\sin^2(\pi p_1/2)}{2h_1^2(2-p_1)p_1}\Big\{-1\nonumber\\&&+2\gamma_E  + \ln4+ \frac{4}{p_1-2}-2\ln\frac{h_1(2-p_1)p_1}{m_s}\nonumber\\&& + \pi\cot\frac{\pi p_1}{2} + 2\ln\sin\frac{\pi p_1}{2}\nonumber\\&& + 2\psi\Big(\frac{p_1}{2}-1\Big)\Big\},
\end{eqnarray}
where $\psi$ is the digamma function and $N(0) = \sqrt{(2-p_1)p_1}$ Then, the ground state energy is given by 
\begin{widetext}
\begin{eqnarray}
\label{e_wh}
{\cal E}(h_1) - {\cal E}(0) &=& -\frac{h_1^2p_1(2-p_1)}{2\pi} - \frac{m_s^2\sin^2(\pi p_1/2)}{4\pi}\Big\{-1+2\gamma_E  + \ln4+ \frac{4}{p_1-2}-2\ln\frac{h_1(2-p_1)p_1}{m_s}\nonumber\\&&+\pi\cot\frac{\pi p_1}{2} + 2\ln\sin\frac{\pi p_1}{2} + 2\psi\Big(\frac{p_1}{2}-1\Big)\Big\}.
\end{eqnarray}
\end{widetext}
The contributions to the ground state energy can be identified as follows. The first term on the right hand side is ${\cal E}_c$, the leading order contribution to the ground state energy when $h_1/M_0, h_2/M_0\rightarrow\infty$. In the second term, the $h_1$-independent term can be identified as the ground state energy ${\cal E}(0)$, while the logarithmic term dependent on $h_1$ is the perturbative correction as we move away from the infinitely large magnetic fields limit. Thus, we get 
\begin{equation}
{\cal E}(h_1) = {\cal E}_c + \delta{\cal E}, 
\end{equation}
where \begin{eqnarray}
\label{e_corr}
\delta{\cal E} = \frac{m_s^2}{2\pi}\sin^2(\pi p_1/2)\ln h_1
\end{eqnarray}
and the ground state energy in the absence of magnetic field as
\begin{eqnarray}
{\cal E}(0) &=& \frac{m_s^2\sin^2(\pi p_1/2)}{4\pi}\Big\{-1+2\gamma_E  + \ln4+ \frac{4}{p_1-2}\nonumber\\&&-2\ln\frac{(2-p_1)p_1}{m_s}+\pi\cot\frac{\pi p_1}{2} \nonumber\\&&+ 2\ln\sin\frac{\pi p_1}{2} + 2\psi\Big(\frac{p_1}{2}-1\Big)\Big\}
\end{eqnarray}
Furthermore, comparing the coefficient of $\ln h_1$ term in Eqs. (\ref{e_cpt}, \ref{e_corr}) leads to an exact expression connecting the interaction parameter $M_0$ of the action with the masses of the $\varphi_1$ solitons: 
\begin{equation}
\label{rel1}
M_0 = m_s\sin\frac{\pi p_1}{2}.
\end{equation}
Here, we have used the definition of $Q_1$ in Eq. [\ref{q1}] and the relation Eq. [\ref{pdef}]. We note that the masses of the $\varphi_2$ solitons are the same as that of the $\varphi_1$ solitons on the integrability manifold. 
\section{The double sine-Gordon model at nonzero-temperature}
\label{TBA}
In this section, we analyze the dSG model at finite temperature. The solitons of this model scatter in a factorized manner. In particular, the scattering of the two quantum numbers of each soliton also occurs independent of each other: $S = S_{p_1}\otimes S_{p_2}$, where on the integrable manifold $p_1+ p_2 = 2$. This relation between $p_1, p_2$ forces one of the amplitude to be in the attractive regime and the other to be in the repulsive regime. This is precisely why the analysis is different from the Fateev model, for which the TBA analysis was done only for the case $p_1, p_2\geq1$~\cite{Fateev1996}. Without loss of generality, we choose $p_1< 1$, i.e. attractive, and $p_2 > 1$ repulsive. In addition to the solitons, the spectrum of the theory also includes $n < 1/p_1$ particles which are neutral with respect to first quantum number, but still carry the second quantum number. They are the bound states of the fundamental `double' solitons and correspond to the ordinary breather poles of the attractive sine-Gordon amplitude of the first sector. The masses of these breathers are given by the standard formula $m_i = 2 m_s\sin[\pi i/2(n+1)]$, where $i = 1,\ldots, n$~\cite{Zamolodchikov1979}. The TBA analysis of this problem is in general highly non-trivial, being each scattering sector source of non diagonally. We will restrict therefore ourselves to the simpler case $p_1 = 1/(n + 1)$ for which the sector one scattering is diagonal. Our TBA analysis will therefore consists in the diagonalization problem of the second sector only, which can be done by means of the Takahashi-Suzuki classification of the solution of the Bethe equations~\cite{Takahashi1972, Fendley1992, Takahashi2005}. 

The periodicity condition for the wave function of this given number of solitons, antisolitons and breathers on a circle of length $L$ becomes the following set of equations: 
\begin{widetext}
\begin{eqnarray}
\prod_{\delta = s,a}\prod_{k_\delta}S^{p_1}_{p\delta}(\theta-\theta_{k_\delta})\prod_{i=1,\ldots, n}\prod_{k_{b_i}}S^{p_1}_{p b_i}(\theta-\theta_{k_{b_i}})\lambda^{p_2}_p(\theta)&=& e^{-i m_p L \sinh\theta},p = s,a, \nonumber\\\prod_{\delta = s,a}\prod_{k_\delta}S^{p_1}_{b_i\delta}(\theta-\theta_{k_\delta})\prod_{j=1,\ldots,n}\prod_{k_{b_j}}S^{p_1}_{b_i b_j}(\theta-\theta_{k_{b_j}})\lambda^{p_2}_{b_i}(\theta)&=& e^{-i m_{b_i} L \sinh\theta},
\end{eqnarray}
\end{widetext}
where $\ i = 1,\ldots, n$. In this equation, $\theta$ is the rapidity of the particle going around the circle, $\theta_{k_s}$, $\theta_{k_a}$, $\theta_{k_{b_i}}$ are the rapidities of the incoming soliton, antisoliton and the breathers respectively. The scattering coefficients for the sine-Gordon model, $S^{p_1}$-s, are well-known~\cite{Zamolodchikov1979}, while $\lambda^{p_2}$ is the contribution of the second phase-shift due to the scattering of the second quantum number. This contribution is computed by diagonalization of the transfer matrix of a second sector soliton going around the world repulsively interacting with solitons and antisolitons with amplitudes $S^{p_2}$. The diagonalization procedure for repulsive sine-Gordon produces in the thermodynamic limit equations for the densities of states in terms of massless pseudoparticles and a physical particle carrying the mass of the soliton. When `gluing' this result with the above equations, the only care will be needed in the identification of the role of the massive physical particles. This is done using the algebraic Bethe ansatz technique~\cite{Korepin1993, Fendley1992, Faddeev1996}. The calculation is nontrivial. First, we do it for $n=2$ and then generalize the results so obtained. The detailed derivation for $n=2$ is given at the end of the manuscript (Appendix \ref{tba_n_2_deriv}) and here, we present directly the final results: 

\begin{eqnarray}
\label{TBA_eq}
\epsilon_p(\theta) &=& \hat\varphi_n \star {\cal L}_{b_n},\ p = s,a, \nonumber\\
\epsilon_{b_n}(\theta) &=& \hat\varphi_n \star ({\cal L}_s + {\cal L}_a + {\cal L}_{b_{n-1}}),\nonumber \\
\epsilon_{b_j}(\theta) &=& \hat\varphi_n \star({\cal L}_{b_{j+1}} + {\cal L}_{b_{j-1}}), \ j = n-1, \ldots, 3,\nonumber\\
\epsilon_{b_2}(\theta) &=& \hat\varphi_n \star({\cal L}_{b_{3}} + {\cal L}_{b_{1}} - {\cal L}_1), \nonumber\\	
\epsilon_{b_1}(\theta) &=& \hat\varphi_n \star({\cal L}_{b_{2}}  - {\cal L}_2),\nonumber \\
\epsilon_{1}(\theta) &=& -m_{b_1}R\cosh\theta + \hat\varphi_n \star({\cal L}_{b_{2}}  - {\cal L}_2),\nonumber\\
\epsilon_{2}(\theta) &=& -\hat\varphi_n \star(L_{b_{1}}  - L_1 + L_3),\nonumber\\
\epsilon_{j}(\theta) &=& -\hat\varphi_n \star (L_{j+1} + L_{j-1}), \ j = n-1, \ldots, 3,\nonumber\\
\epsilon_n(\theta) &=& -\hat\varphi_n \star(L_{n-1} + L_{n+1} + L_{n+2}), \nonumber\\
\epsilon_{k}(\theta) &=& -\hat\varphi_n \star L_{n},\ k = n+1, n+2.
\end{eqnarray}

In the above equation, the TBA kernel is given by $\hat\varphi_n(\theta) = (n+1)/\cosh[(n+1)\theta]$, ${\cal L}(\theta) = \ln[1 + e^{\epsilon(\theta)}]$, and $L(\theta) = \ln[1 + e^{-\epsilon(\theta)}]$. Finally, we use the following convention for the convolution: $a \star b = \int \frac{d\theta'}{2\pi} a(\theta-\theta') b(\theta')$. The pictorial representation for the TBA equations is given Fig. \ref{TBA_pic}. The solid circles denote the physical massive particles, while the empty ones denote the pseudoparticles. The cross on the empty circle encodes which of the  pseudoparticles has a massive term in its TBA equation. The connectivity of the diagram encodes the structure of the TBA equations. Finally, the arrows denote, for the TBA equation for a given particle, the sign of the contribution of the different particles. 
\begin{figure}
\centering
\includegraphics[width = 0.48\textwidth]{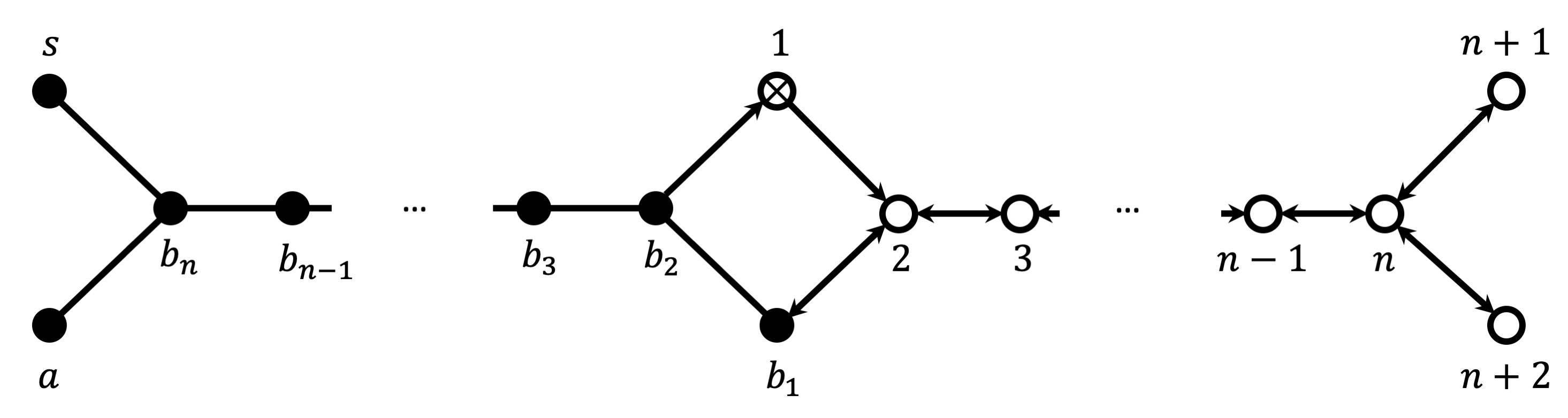}
\caption{\label{TBA_pic}TBA diagram for the dSG model. The solid circles denote the physical massive particles: soliton ($s$), antisoliton($a$) and breathers ($b_j$, $j = 1,\ldots, n$). The empty circles denote pseudoparticles ($j=1,\ldots, n+2$) needed to diagonalize the dSG scattering matrix. The pseudoparticle $1$ has a cross on it, which indicates that this particle has a mass term in its TBA equation. As usual, the connectivity of the diagram encodes which particles show up in the  TBA equation of a given particle. The arrows on the links encode the sign of the term on the right hand side of the TBA equation. For instance, for a link connecting particles $p, q$, if there is an arrow incident on $q$ and none on $p$, then the right hand side of the TBA equation for $p$ has a minus sign, while the same for $q$ has a plus sign. }
\end{figure}
Note the somewhat special structure of the TBA equations. In contrast to the usual structure of the TBA equations, in Eq. [\ref{TBA_eq}], only one pseudoparticle ($1$) has a (negative!) mass term, while the physical massive particles do not. This is because the equations are written in terms of both the functions ${\cal L}$-s and $L$-s, which allows us to write the TBA equations concisely in terms of a single universal kernel $\hat\varphi_n$. When written only in terms of the $L$-s, the TBA equations are in the `standard form' and do have the expected structure. 

Note the difference of the TBA equations from those obtained for the Fateev model. The latter model involves in addition to the two bosonic fields considered in this work, a noncompact bosonic field $\varphi$ with coupling $\beta$. The TBA for the Fateev model has been done in the regime when both the couplings $\alpha_1, \alpha_2$ are larger than $\sqrt{2\pi}$. In this regime, both the sine-Gordon models are in the repulsive regime (without any breathers in either sector) and the resultant TBA is obtained by `gluing' two sine-Gordon TBA diagrams in the repulsive regime. In contrast, in the dSG model, the pure quantum integrable manifold corresponds to $\alpha_1^2 + \alpha_2^2 = 4\pi$, with $\alpha_1\neq\alpha_2$. Thus, the breathers in one of the sectors have to be taken into account to correctly obtain the thermodynamics of the dSG model. This difference is visible by taking the naïve limit of $\beta = 0$ in Eq. 33 of Ref. \onlinecite{Fateev1996}, which gives divergent results for the ground state energy. 

To check the TBA equations, we first compute the central charge in the UV limit. The central charge is given by 
\begin{equation}\label{cUV1}
c_{\rm{UV}} = \frac{6}{\pi^2}\int d\theta\cosh\theta\Big[L_s(\theta) + L_a(\theta) + \sum_{j=1}^n\frac{m_{b_i}}{m_s}L_{b_i}(\theta)\Big]
\end{equation}
For TBA equations with off-diagonal scattering, the central charge can be computed in the usual manner using the solutions of the energies $\epsilon_s$, $\epsilon_a$, $\epsilon_{b_i}$, $\epsilon_i$-s in the UV ($T\rightarrow\infty$) and IR ($T\rightarrow0$) limits~\cite{Zamolodchikov1991a, Zamolodchikov1991b}. The resultant expression for the central charge can be written as 
\begin{widetext}
\begin{eqnarray}
\label{cUV}
c_{\rm{UV}} &=& \frac{6}{\pi^2}\Big\{2{\cal L}_{\rm{dlog}}\Big(\frac{x_s}{1+x_s}\Big) + \sum_{j=2}^n\Big[{\cal L}_{\rm{dlog}}\Big(\frac{x_j}{1+x_j}\Big) + {\cal L}_{\rm{dlog}}\Big(\frac{x_{b_j}}{1+x_{b_j}}\Big)\Big]\nonumber\\&& +2 {\cal L}_{\rm{dlog}}\Big(\frac{x_{n+1}}{1+x_{n+1}}\Big)+ {\cal L}_{\rm{dlog}}\Big(\frac{x_{b_1}}{1+x_{b_1}}\Big)+ {\cal L}_{\rm{dlog}}\Big(\frac{1}{1+x_{1}}\Big)\nonumber\\&&-\sum_{j=2}^n{\cal L}_{\rm{dlog}}\Big(\frac{y_j}{1+y_j}\Big) - {\cal L}_{\rm{dlog}}\Big(\frac{1}{1+y_{1}}\Big)- 2 {\cal L}_{\rm{dlog}}\Big(\frac{y_{n+1}}{1+y_{n+1}}\Big)\Big\},
\end{eqnarray}
\end{widetext}
where ${\cal L}_{\rm{dlog}}$ is the Rogers dilogarithm function and we have used $\epsilon_s = \epsilon_a$, $\epsilon_{n+1} = \epsilon_{n+2}$. Furthermore, $x_p=e^{-\epsilon_p}$, $p = s,a,b_i, i$, $i=1,\ldots,n$ in the UV limits, while the $y_p=e^{-\epsilon_p}$ in the IR limit. Note that in the IR limit, the energies of the massive physical particles diverge and so, they do not contribute to the expression of the central charge~\cite{Zamolodchikov1991a}. The solutions of the TBA equations in the UV limit are given by
\begin{eqnarray}
x_{n+1} &=&\frac{1}{x_s} = n, \ x_1 = x_{b_1} = 1,\nonumber\\ x_j &=& \frac{1}{x_{b_j}}=j^2-1, \ j=2, \ldots, n, 
\end{eqnarray}
while for the IR limit, the non-zero $y$-s are given by
\begin{eqnarray}
y_{n+1} &=& n-\frac{1}{2},\ y_1 = 3,\nonumber\\ y_j &=& \Big(j-\frac{1}{2}\Big)^2-1, j = 2, \ldots, n. 
\end{eqnarray}
Plugging these solutions in Eq. [\ref{cUV}] and using some remarkable dilogarithm identities~\cite{Lewin1981}, we arrive the desired result of 
\begin{equation}
c_{\rm{UV}} = 2.
\end{equation}
Next, we compute the effective central charge as we move away from the UV limit. The perturbation ${\cal S}_{\rm{int}}$ has dimension $(\alpha_1^2 + \alpha_2^2)/(4\pi)$. Thus, on the integrable manifold, the perturbation has dimension 1. This leads to logarithmic corrections to the central charge~\cite{Zamolodchikov1991a, Fendley1993} as one moves away from the UV limit, in addition to the polynomial corrections in powers of $(m_sR)^2$~\cite{Klassen1991}. We compute these corrections by solving the TBA equations numerically and doing a numerical fit of the resulting expression for the central charge. The coefficient of the logarithmic correction for the different cases is then verified using CPT. These calculations are described below for $n=2$ and $n=3$. 

\begin{figure}
\centering
\includegraphics[width = 0.48\textwidth]{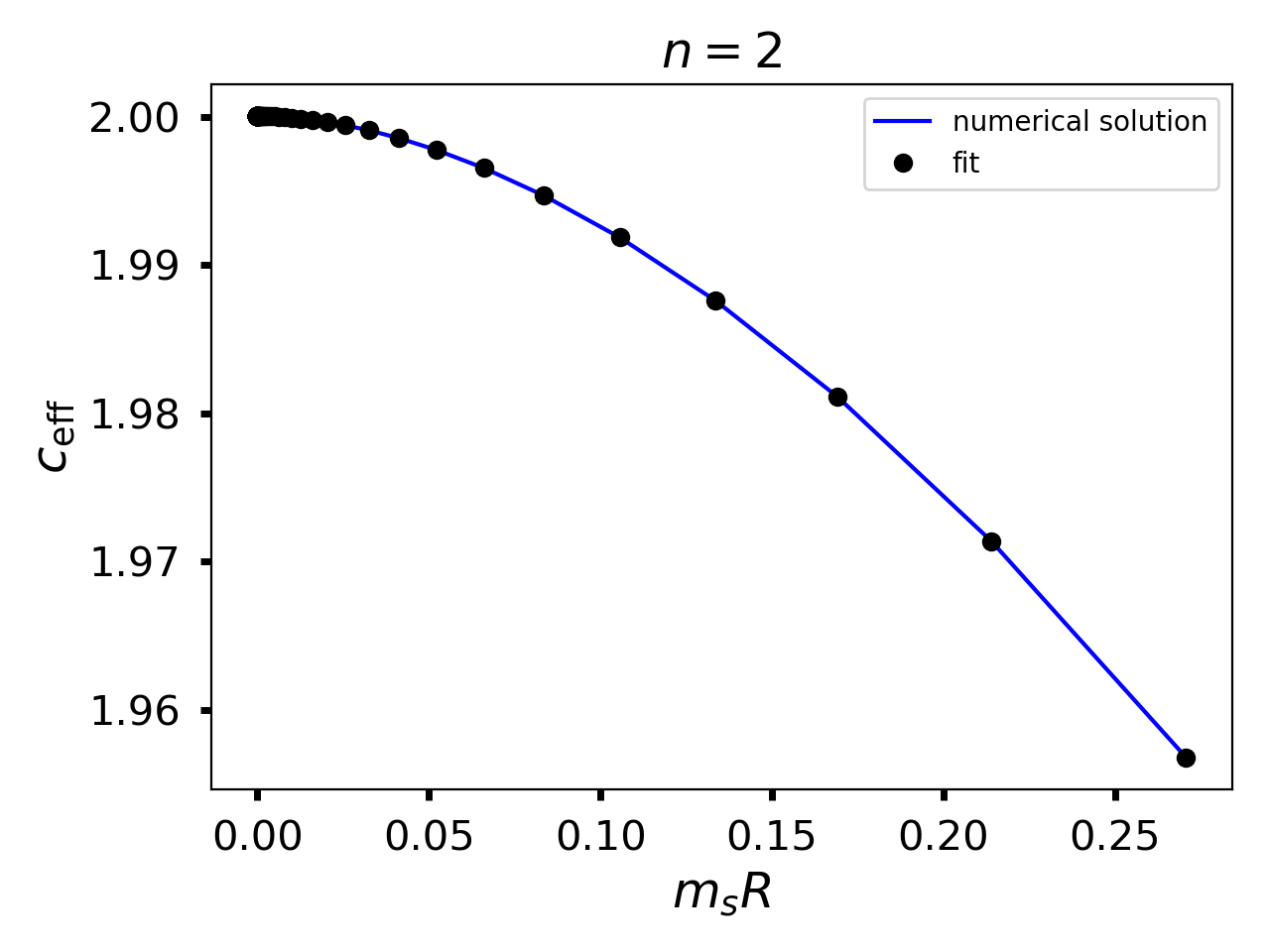}
\caption{\label{cUVn_2} Effective central charge for $n=2$. The solid blue line is obtained by solving the TBA equations given by Eq. [\ref{TBA_n_2}]. The black dots are obtained by performing a fit to Eq. [\ref{cUV_n}]. }
\end{figure}

Consider the case $n=2$, $i.e.,$ $p_1 = 1/3, p_2 = 5/3$. The first sector spectrum consists of soliton $s$, antisoliton $a$ and two breathers $b_1$ and $b_2$ whose masses are $m_{b_1} = m_s$ and $m_{b_2} = \sqrt{3}m_s$. For the numerical solutions of the TBA equations, it is convenient to use the equations in the standard form, given by
\begin{widetext}
\begin{eqnarray}
\label{TBA_n_2}
\epsilon_p(\theta) &=& m_p R \cosh\theta + \hat\phi_1\star (L_s +L_a+ L_{b_1} - L_1) + \hat\phi_3\star L_{b_2},\ p = s,a, \nonumber\\
\epsilon_{b_2}(\theta)  &=&m_{b_2}R\cosh\theta + \hat\phi_3\star (L_s+L_a + L_{b_1} - L_1) + 2\hat\phi_1\star L_{b_2}\nonumber\\
\epsilon_{b_1}(\theta)  &=&m_{b_1}R\cosh\theta + \hat\phi_1\star (L_s +L_a+ L_{b_1} - L_1) + \hat\phi_3\star L_{b_2}- \hat\phi_2\star L_{2}, \nonumber\\
\epsilon_{1}(\theta)  &=&\hat\phi_1\star (L_s + L_a+L_{b_1} - L_1) + \hat\phi_3\star L_{b_2}- \hat\phi_2\star L_{2}, \nonumber\\
\epsilon_{2}(\theta)  &=&-\hat\phi_2\star (L_{b_1} - L_1 +L_2+L_3), \nonumber\\
\epsilon_{k}(\theta)  &=&-\hat\phi_2\star L_2, \ k=3,4,
\end{eqnarray}
\end{widetext}
where the kernels are given by 
\begin{eqnarray}
\hat\phi_1(\theta) &=& \frac{3}{2\cosh(3\theta/2)},\ \hat\phi_2(\theta) = \frac{3}{\cosh(3\theta)}, \nonumber\\ \hat\phi_3(\theta) &=& 3\sqrt{2}\frac{\cosh(3\theta/2)}{\cosh(3\theta)}.
\end{eqnarray}
We solve Eq. [\ref{TBA_n_2}] numerically and plug it in Eq. [\ref{cUV1}] to obtain the effective central charge $c_{\rm{eff}}$. The result is shown in Fig. \ref{cUVn_2}.  
The first six correction terms capturing the deviation from the UV limit is obtained by fitting the numerical data to the following expression:
\begin{eqnarray}
\label{cUV_n}
c_{\rm{eff}} &=& c_0 + a^{(2)}_{\rm{log}}\frac{(m_sR)^2}{(2\pi)^2}\ln (m_sR)\nonumber\\&& + \sum_{j=1}^5 a^{(2)}_j(m_sR)^{2j},  
\end{eqnarray}
where the superscript $2$ denotes the value of $n$. This leads to \begin{eqnarray}\label{cUV_n_2}
c_0 &=&2, \ a^{(2)}_{\rm{log}} = 6, 
\ a^{(2)}_1 = -0.39283, \ a^{(2)}_2 = 0.01051, \nonumber\\\ a^{(2)}_3 &=& -0.01554,\ a^{(2)}_4 = 0.25742, \ a^{(2)}_5 = -1.54212. \nonumber
\end{eqnarray}
Identical calculations can be done for $n=3$. We only provide the final fit to the effective central charge:
 \begin{eqnarray}\label{cUV_n_3}
c_0 &=&2, \ a^{(3)}_{\rm{log}} = 3.5149, 
\ a^{(3)}_1 = -0.29775, \ a^{(3)}_2 = 0.00739, \nonumber\\\ a^{(3)}_3 &=& -0.00818,\ a^{(3)}_4 = 0.1851, \ a^{(3)}_5 = -1.2709. \nonumber
\end{eqnarray}
The coefficient of the logarithmic term in the effective central charge calculation can be checked analytically using CPT. Taking Eq. [\ref{eqcpt}] for $h_1 =0$ and using Eq. [\ref{rel1}], we arrive at the expression for the free energy:
\begin{eqnarray}
f  &=& -\frac{1}{LR}\ln Z= -\frac{\pi c_{\rm{eff}}}{6R^2}\nonumber\\&=& -\frac{1}{LR}\ln Z_0 - \frac{m_s^2}{\pi}\sin^2\frac{\pi}{2(n+1)}\ln\Big(\frac{R}{2a}\Big). 
\end{eqnarray}
Thus, we arrive at an analytical expression for the logarithmic correction in $c_{\rm{eff}}$, proportional to  $\frac{(m_sR)^2}{(2\pi)^2}\ln (m_sR)$, denoted by $a_{\rm{log}}$:
\begin{equation}
a_{\rm{log}}=24\sin^2\frac{\pi}{2(n+1)}.
\end{equation}
For $n=2$ and $n=3$, this coefficient is given by $6$ and $3.5147$, which are in good agreement with what was obtained by numerically solving the TBA equations (Eqs. [\ref{cUV_n_2}, \ref{cUV_n_3}]). We have checked our results for $n=4$ and $n=5$ as well. 

\section{The double sine-Gordon model with quantum circuits}
\label{qcirc}
In this section of the appendix, we present some additional details on the realization of the dSG with quantum electronic circuits. As mentioned in the main text, the dSG model can be viewed as two sine-Gordon (SG) models coupled by a mutual inductance and a capacitance. Below, we sketch how the different terms of the circuit contribute to the different terms of the action presented in Eq. (1) of the main text. 

\begin{figure}
\centering
\includegraphics[width = 0.48\textwidth]{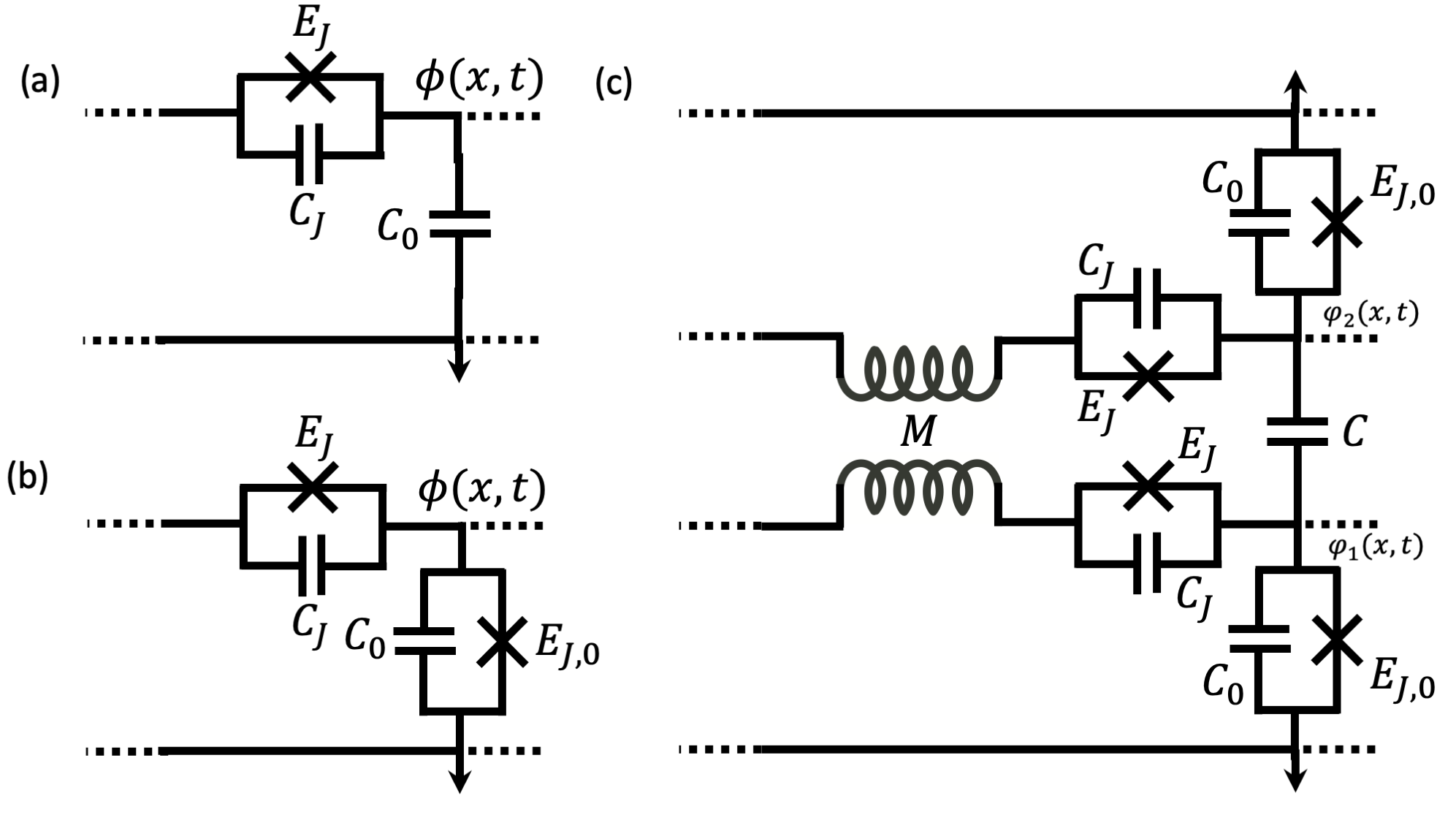}
\caption{\label{circuit} Unit cells for a free-boson quantum field theory [panel (a)], the quantum sine-Gordon model [panel (b)] and the double sine-Gordon model [panel (c)].  }
\end{figure}

Consider the circuit in Fig. \ref{circuit} (a). Define the node-flux at a point in space-time $(x,t)$ as $\phi(x,t)$~\cite{Devoret_1997}, which plays the role of the bosonic field. As shown in Refs. \onlinecite{Clerk_Schoelkopf_2010, Goldstein2013, Roy2016, Roy2018a}, a 1D array built out of this unit cell, in the continuum limit, provides a circuit realization of a 1D Luttinger liquid. The action of the latter is given by ${\cal S}_{\rm{LL}} = \frac{1}{2\pi K}\int d^2x (\partial_\mu\phi)^2$, where we have rescaled the space and time axis to set the plasmon velocity to be unity. It is essential that $E_J, E_{C_0}\gg E_{C_J}$ so that the phase-slip rates are exponentially suppressed. This ensures that despite the presence of the nonlinear Josephson element on the horizontal links, the resultant action is Gaussian.  Addition of a Josephson junction on vertical links, adds the Josephson cosine-potential to the action, leading to the sine-Gordon action, where ${\cal S}_{\rm{SG}} ={\cal S}_{\rm{LL}} -E_{J_0}\int d^2x\cos\phi$, where $E_{J_0}$ is the strength of the Josephson nonlinearity. It is this nonlinearity that gives rise to the cosine nonlinearity of the SG action. In the quantum field theory notation, the SG coupling constant $\beta$ is given by $\beta = \sqrt{4\pi K}$. The free-fermion point of the SG model corresponds to $\beta = \sqrt{4\pi}$, while $\beta\leq(\geq)\sqrt{4\pi}$ corresponds to the attractive (repulsive) regime of the model. 

From the above circuits, it is straightforward to infer the circuit realizing the dSG action, presented in Fig. \ref{circuit} (c). Consider the upper and lower halves of the array, without the mutual inductance $M$ and the capacitance $C$, which couples the two halves. From the above explanation, it is clear that each of the two halves of the array realize the SG actions, given by ${\cal S}_{{\rm{SG}}, i}$, $i = 1,2$: 
\begin{eqnarray}
{\cal S}_{{\rm{SG}}, i} &=& \frac{1}{2\pi K}\int d^2x (\partial_\mu\varphi_i)^2\nonumber\\&& -E_{J_0}\int d^2x \cos\varphi_i, \ i=1,2.
\end{eqnarray}
Now, consider the mutual inductance $M$ and the capacitance $C$. The first leads to a term in the action
\begin{equation}
{\cal S}_M = -M\int d^2x \partial_x \varphi_1\partial_x \varphi_2,
\end{equation}
while the second leads to 
\begin{equation}
{\cal S}_C = \frac{C}{2}\int d^2x (\partial_t \varphi_1-\partial_t \varphi_2)^2. 
\end{equation}
The last two equations follow directly from the current-voltage constitutive relations of the circuit elements under consideration. The total action of the dSG circuit is given by ${\cal S}_{{\rm{SG}}, 1} + {\cal S}_{{\rm{SG}}, 2}$ $+ {\cal S}_M + {\cal S}_C$. From this action, simple algebraic manipulations give rise to the dSG action in terms of the fields $\varphi_{1,2}$, given by
\begin{eqnarray}
{\cal S}_{\rm{dSG}} &=& \frac{A}{2}\int d^2x[(\partial_\mu\varphi_1)^2 + (\partial_\mu\varphi_2)^2]  + B\int d^2x \partial_\mu \varphi_1\partial_\mu\varphi_2\nonumber\\&& - \frac{M_0}{\pi}\int d^2x\{\cos(2\sqrt{\pi}\varphi_1) + \cos(2\sqrt{\pi}\varphi_2)\},
\end{eqnarray}
where $A = \pi(1/\alpha_1^2 + 1/\alpha_2^2)$ and $B = \pi(1/\alpha_1^2-1/\alpha_2^2)$. In terms of the circuit components, 
\begin{equation}
\alpha_1^2 = \frac{K\pi}{2}, \ \alpha_2^2 = \frac{\pi}{2/K + 4\pi C}, 
\end{equation}
which are the expressions provided in the main text of the article. 

In the above derivation of the dSG action from the circuit, we have implicitly assumed that the unit cell is repeated an infinite number of times with periodic/open boundary conditions. However, real experiments are done with finite number of such unit cells. There are two main difficulties in realizing arrays very large number of junctions. First, in the derivation of the dSG action, we have neglected the phase-slips in the Josephson junctions in the horizontal links. However, for a large enough number of unit cells, this rate is no longer negligible. A crude estimate for the maximum number of unit cells is provided by the inverse of the phase-slip rate and is given by $e^{\sqrt{E_J/E_{C_J}}}$~\cite{Manucharyan2009, Goldstein2013}. The same problem also provides a limitations on the frequencies over width the phase-slip rates can be safely neglected and is given by $u\sqrt{C_J/C_0}/a_0$, where $a_0$ is the lattice spacing and $u=a_0\sqrt{2E_{C_0}E_J}$, the plasmon velocity~\cite{Goldstein2013}. Second, the superinductance of the Josephson junction on the horizontal links should not be shunted by the capacitance to ground. This is ensured by choosing $N<C_J/C_0$~\cite{Manucharyan2009}. Despite these difficulties, arrays with up to ${\cal O}(10^5)$ Josephson junctions have been reliably built and experimentally analyzed~\cite{Kuzmin2018} and we are optimistic that our analytical predictions can be experimentally verified. 

Next, we comment on some of the possible measurements in the circuit. Estimation of $\alpha_1,\alpha_2$ can be done by biasing with two current sources (the `magnetic fields' $h_1,h_2$ considered earlier)  and measuring the transmission of an input field through the dSG lattice. This provides an estimate of the average current, $\langle\partial_x\varphi_{1,2}\rangle$, flowing through the horizontal links in the circuit. A different approach will be to locally probe the dSG circuit by coupling resonators to a unit cell and measuring current/voltage correlations. While transport properties can be measured in the ways mentioned above, the specific heat can  be measured using AC calorimetry~\cite{Fominaya1997}. The latter method involves inducing temperature oscillations on the array patterned on a membrane. The amplitude of the oscillations, together with the applied amplitude and frequency of the applied AC power provides the specific heat~\cite{Fominaya1997, Rabani2008}.  A detailed theoretical analysis for the specific heat measurements for the sine-Gordon model has been made in Refs.  ~\onlinecite{Guarcello2018, Guarcello2018a}. 

We note that opening the circuit to drive and dissipation leads to an interesting and physically relevant problem of dissipative quantum field theories, a largely open field. The  calculations made for the sine-Gordon model~\cite{Hida1984, Hida1985} indicate a finite width for the sine-Gordon minima which presumably leads to a finite width for the different particles in the spectrum. We speculate that a similar phenomenon may occur in the dSG model. However, a full quantum analysis is still an open problem even for the sine-Gordon model and we hope to return to this problem in the future. 

\section{Derivation of the TBA equations for $n=2$}
\label{tba_n_2_deriv}
Consider again the case $n=2$, $i.e.,$ $p_1 = 1/3, p_2 = 5/3$. The first sector spectrum consists of soliton $s$, antisoliton $a$ and two breathers $b_1$ and $b_2$ whose masses are $m_{b_1} = m_s$ and $m_{b_2} = \sqrt{3}m_s$. We refer to the two sine-Gordon sectors as $T_1, T_2$ which correspond to the couplings $\alpha_1, \alpha_2$. 
The periodicity condition for the wave function of a given number of solitons, 
antisolitons and breathers on a circle of length $L$ becomes the 
following set of equations
\begin{widetext}
\begin{eqnarray}
\prod_{k_s} S^{p_1}_{s s}(\theta-\theta_{k_s})
\prod_{k_a} S^{p_1}_{s a}(\theta-\theta_{k_a}) 
\prod_{k_{b_1}} S^{p_1}_{s b_1}(\theta-\theta_{k_{b_1}}) 
\prod_{k_{b_2}} S^{p_1}_{s b_2}(\theta-\theta_{k_{b_2}}) 
\lambda^{p_2}_s (\theta | \{\theta_{k_s}\},\{\theta_{k_a}\},
\{\theta_{k_{b_1}}\}, \{\theta_{k_{b_2}}\}) &=& e^{-i m_s L \sinh\theta}
\nonumber \\
\prod_{k_s} S^{p_1}_{b_1 s}(\theta-\theta_{k_s})
\prod_{k_a} S^{p_1}_{b_1 a}(\theta-\theta_{k_a}) 
\prod_{k_{b_1}} S^{p_1}_{b_1 b_1}(\theta-\theta_{k_{b_1}}) 
\prod_{k_{b_2}} S^{p_1}_{b_1 b_2}(\theta-\theta_{k_{b_2}}) 
\lambda^{p_2}_{b_1} (\theta | \{\theta_{k_s}\},\{\theta_{k_a}\},
\{\theta_{k_{b_1}}\}, \{\theta_{k_{b_2}}\}) &=& e^{-i m_1 L \sinh\theta} 
\nonumber \\
\prod_{k_s} S^{p_1}_{b_2 s}(\theta-\theta_{k_s})
\prod_{k_a} S^{p_1}_{b_2 a}(\theta-\theta_{k_a}) 
\prod_{k_{b_1}} S^{p_1}_{b_2 b_1}(\theta-\theta_{k_{b_1}}) 
\prod_{k_{b_2}} S^{p_1}_{b_2 b_2}(\theta-\theta_{k_{b_2}}) 
\lambda^{p_2}_{b_2} (\theta | \{\theta_{k_s}\},\{\theta_{k_a}\},
\{\theta_{k_{b_1}}\}, \{\theta_{k_{b_2}}\}) &=& e^{-i m_2 L \sinh\theta}  
\nonumber \\
&& \label{periodicity}
\end{eqnarray}
\end{widetext}
where $\theta$ is the rapidity of the particle going around the 
world, $\theta_{k_{s,a,b_1,b_2}}$ the rapidities of the external ``incoming'' 
solitons, antisolitons and breathers and 
in which $\lambda^{p_2}_i$ represents the contribution to the phase shift
of the scattering of the second quantum number.
We compute these contributions by the diagonalizing  the transfer 
matrix of a $T_2$-soliton going around the world repulsively interacting  
with solitons and antisolitons with amplitudes $S^{p_2}_{SG}$.
The diagonalization procedure for repulsive sine-Gordon produces 
in the thermodynamic limit equations for the densities of states in terms of 
massless pseudoparticles and a physical particle carrying the mass 
of the soliton. When ``gluing'' this result with the equations (\ref{periodicity})
the only care will be needed in the identification of the role 
of the massive physical particles.
Through the algebraic Bethe ansatz technique~\cite{Zamolodchikov1991a, Fendley1992}, 
the eigenvalues of the repulsive sine-Gordon transfer matrix are
\begin{widetext}
\begin{equation}
\lambda^{p_2}(\theta,\{\theta_i\},\{y_r\}) = 
\prod_i S^{p_2}_{s s}(\theta - \theta_i)
\prod_r \frac{\sinh\frac{1}{p_2}(i\pi -y_r -\theta)}
{\sinh\frac{1}{p_2}(y_r -\theta)}\,\,\,\,\,\,.
\label{betheeigen}
\end{equation}
\end{widetext}
The rapidities $y_r$ label the Bethe eigenvectors of the transfer matrix
and have to satisfy the Bethe equation

\begin{eqnarray}
\prod_i \frac{\sinh\frac{1}{p_2}(i\pi -y_r -\theta)}
{\sinh\frac{1}{p_2}(y_r -\theta)} =
- \prod_{r'} \frac{\sinh\frac{1}{p_2}(y_r - y_{r'} - i\pi)}
{\sinh\frac{1}{p_2}(y_r - y_{r'} + i\pi)}\,\,\,\,\,\,.
\label{betheeq}
\end{eqnarray}

The presence of the external particles does not invalidate the string
classification \cite{Takahashi1972, Takahashi2005} of the rapidities $y_r$ in the thermodynamic limit.
This classification applies actually to the case $p_2 >2$.
It is easy to convince yourself that the classification of the $y_r$'s
of equation (\ref{betheeq}) is equivalent to the classification of the
$y'_r = (1- p_2) y_r$ of an equation like (\ref{betheeq}) but with 
$p'_2 = (1- 1/p_2)^{-1}$.
For the general case $p_2 = 2 - 1/(n+1)$ we have the following pattern
of $n+2$ strings $y^{(j)}$ 
\begin{widetext}
\begin{equation}
y^{(j)}_k = \Re{y^{(j)}} - \frac{i\pi}{2}\frac{n}{n+1} (2 j -1 -2 k) -
\frac{i\pi}{2}p_2 \frac{1-v_j}{2}\,\,,\,\,\, k= 1,2, \dots N_j\,\,,
\,\,\,\,j=1,\ldots,n+2\,\,.
\label{strings}
\end{equation}
The lengths $N_j$ and the parities $v_j$ of the string $y^{(j)}$ are
\begin{eqnarray}
N_1 = 1\,\,; \,\,\,\,\,\, N_j = 2j - 3 \,\,, \,\,\,j=2,3,\ldots,n+1\,\,;
&&N_{n+2}=2 \nonumber\\
v_1 = 1\,\,;  \,\,\,\,\,\,v_j = (-1)^{j-1} \,\,, \,\,\,j=2,3,\ldots,n+1\,\,;
&&v_{n+2}=1\,\,\,\,\,\,.
\label{parities}
\end{eqnarray}
\end{widetext}
Since there cannot exist different strings with the same length and parity
it is useful to denote the strings with $N_v$. 
The center of a string, i.e. its real part, plays the role 
in the thermodynamical analysis of the rapidity of a massless pseudoparticle
$N_v$ to which we can associate a density $P_{N_v}$ of states containing it 
and a density of occupied states $P^+_{N_v}$.
As usual the Yang-Yang relations between the densities is obtained
by plugging the classification (\ref{strings}) in the Bethe equation 
(\ref{betheeq}) and by taking the imaginary part of the 
logarithmic derivative with respect to the external rapidity.
The result for the pseudoparticles is
\begin{equation}
P_i(y) = (-1)^{v_i}\left\{\hat\phi_{i 0}\star P^+_0 +
\sum_j \hat\phi_{i j} \star P^+_j\right\}
\label{sgdensities}
\end{equation}
where 
\begin{widetext}
\begin{eqnarray}
\hat\phi_{i j}(y) &=&-i d/dy\log\left\{\prod_{k_i = 1}^{N_i} \prod_{k_j = 1}^{N_j} 
\frac{\sinh\frac{1}{p_2}[y-\frac{i\pi}{2}\frac{n}{n+1} 
(N_i-N_j -2k_i+2k_j) -\frac{i\pi}{2} p_2 \frac{1-v_i v_j}{2}- i\pi ]}
{\sinh\frac{1}{p_2}[y-\frac{i\pi}{2}\frac{n}{n+1} 
(N_i-N_j -2k_i+2k_j)\frac{i\pi}{2}p_2-\frac{1-v_i v_j}{2}+i\pi ]}\right\}
\nonumber\\
&=&
 (-1)^{v_i v_j} \frac{2}{p_2}\sum_{k_i =1}^{N_i}\sum_{k_j =1}^{N_j}
\frac{\sin\frac{\pi}{p_2}[\frac{n}{n+1}(N_i-N_j -2k_i+2k_j)+2]}
{\cosh\frac{2 y}{p_2} - (-1)^{v_i v_j}
\cos\frac{\pi}{p_2}[\frac{n}{n+1}(N_i-N_j -2k_i+2k_j)+2] },\\
\hat\phi_{i 0}(y) &=&-i d/dy\log\left\{\prod_{k_i = 1}^{N_i} 
\frac{-\sinh\frac{1}{p_2}[y-\frac{i\pi}{2}\frac{n}{n+1} 
(N_i+2 -2k_i) -\frac{i\pi}{2} p_2 \frac{1-v_i}{2}]}
{\sinh\frac{1}{p_2}[y-\frac{i\pi}{2}\frac{n}{n+1} 
(N_i+2 -2k_i)\frac{i\pi}{2}p_2-\frac{1-v_i}{2}-i\pi ]}\right\}
\nonumber\\
&=&
(-1)^{v_i} \frac{2}{p_2}\sum_{k_i =1}^{N_i}
\frac{\sin\frac{\pi}{p_2}[\frac{n}{n+1}(N_i + 2 -2k_i)]}
{\cosh\frac{2 y}{p_2} - (-1)^{v_i}
\cos\frac{\pi}{p_2}[\frac{n}{n+1}(N_i +2 -2k_i)] }
\end{eqnarray}
\end{widetext}
are the kernels of the particle $i$ with the physical particle
denoted with $0$ and with the other pseudoparticles $j$.
The convolution normalization is 
$a\star b = \int\frac{d\theta'}{2\pi} a(\theta-\theta') b(\theta')$.
For sine-Gordon the method would go on with the equation for the density
of the physical particle state, obtainable from the periodicity equation.
But in the case we are examining now the role of the physical particle
is played by the $T_1$ soliton, antisoliton and breathers.
Therefore in our case we have to modify the equation (\ref{sgdensities})
to the equation
\begin{widetext}
\begin{equation}
P_i(y) = (-1)^{v_i}\left\{\phi_{i 0}\star (P^+_s + P^+_a) +
\sum_j \hat\phi_{i b_j} \star P^+_{b_j}+
\sum_j \hat\phi_{i j} \star P^+_j\right\}
\label{densities}
\end{equation}
where the breather kernels are
\begin{equation}
\hat\phi_{i b_j}(\theta) = \hat\phi_{i 0}\left(\theta + 
\frac{i\pi}{2}\frac{n-j+1}{n+1}\right)+
\hat\phi_{i 0}\left(\theta  -\frac{i\pi}{2}\frac{n-j+1}{n+1}\right).
\end{equation}
\end{widetext}
The equations for the densities of the physical particles states can be 
obtained by taking the logarithmic derivative of equations (\ref{periodicity})
with the explicit formulas for the eigenvalues (\ref{betheeigen}) and
(\ref{strings}). The result is
\begin{widetext}
\begin{eqnarray}
P_s(\theta) &=& \frac{m_s L}{2\pi} \cosh\theta +
\hat\phi_{s s}\star (P_s^+ +P_a^+) +\sum_j \hat\phi_{s b_j}\star P^+_{b_j}
+\sum_j \hat\phi_{s j}\star P^+_j \nonumber\\
P_{b_i}(\theta) &=& \frac{m_i L}{2\pi} \cosh\theta +
\hat\phi_{b_i s}\star (P_s^+ +P_a^+) +\sum_j \hat\phi_{b_i b_j}\star P^+_{b_j}
+\sum_j \hat\phi_{b_i j}\star P^+_j \,\,\,\,.
\end{eqnarray}
\end{widetext}
The new kernels are 
\begin{widetext}
\begin{eqnarray}
\hat\phi_{s s}(\theta)&=&-i\frac{d}{d\theta} \log S^{p_1}_{s s}(\theta)
+\phi_{0 0}(\theta) \nonumber\\
\hat\phi_{s b_j}(\theta)&=& -i\frac{d}{d\theta} \log S^{p_1}_{s b_j}(\theta)
+ \left[\phi_{0 0}\left(\theta +\frac{i\pi}{2}\frac{n-j+1}{n+1}\right) 
+\hat\phi_{0 0}\left(\theta -\frac{i\pi}{2}\frac{n-j+1}{n+1}\right)\right] 
\nonumber\\
\hat\phi_{s i}(\theta)&=&\hat\phi_{i 0}(\theta)\nonumber\\
\hat\phi_{b_i s}(\theta)&=&\hat\phi_{s b_i}(\theta)\nonumber\\
\hat\phi_{b_i b_j}(\theta)&=&
-i\frac{d}{d\theta} \log S^{p_1}_{b_i b_j}(\theta)+
\left[\hat\phi_{0 0}\left(\theta+ \frac{i\pi}{2}\frac{2n-i-j+2}{n+1}\right) 
+\hat\phi_{0 0}\left(\theta+\frac{i\pi}{2}\frac{i-j}{n+1}\right) 
\right. \nonumber\\
& &\left.
+\hat\phi_{0 0}\left(\theta-\frac{i\pi}{2}\frac{i-j}{n+1}\right) 
+\hat\phi_{0 0}\left(\theta-\frac{i\pi}{2}\frac{2n-i-j+2}{n+1}\right)
\right]\nonumber\\
\hat\phi_{b_i j}(\theta)&=&\hat\phi_{j b_i}(\theta)
\end{eqnarray}
\end{widetext}
where we denote with
\begin{eqnarray}
\hat\phi_{0 0}(\theta) &=&-i\frac{d}{d\theta} \log S^{p_2}_{s s}(\theta) 
\nonumber\\
&=& \int dk e^{ik\theta}\frac{\sinh(p_2 -1)\frac{\pi k}{2}}
{2 \sinh p_2\frac{\pi k}{2} \cosh\frac{\pi k}{2}}
\end{eqnarray}
the soliton-soliton kernel of the pure $p_2$-sineGordon TBA.
In order to be concrete, let's come back to the specific example
$p_1=1/3$ and $p_2=5/3$.
The equations for the density can be written in Fourier space with
$\hat\phi(\theta)=\int dk e^{ik\theta}\tilde{\phi}(k)$ and assume the form
\begin{widetext}
\begin{eqnarray}
{\cal FT}\left[P_s-\frac{m_s L}{2\pi}\cosh\theta\right]&=&
-\frac{2 c_{1/6} s_{1/3}}{s_{5/6}} (\tilde{P}^+_s +\tilde{P}^+_a)
-\frac{ s_{1/2}}{s_{5/6}} \tilde{P}^+_{b_1}
-\frac{4 c_{1/6}^2 s_{1/3}}{s_{5/6}} \tilde{P}^+_{b_2}\nonumber\\
&&
+\frac{ s_{1/2}}{s_{5/6}} \tilde{P}^+_1
-\frac{ s_{1/3}}{s_{5/6}} \tilde{P}^+_{1_-}
+\frac{ s_{1/6}}{s_{5/6}} \tilde{P}^+_2
-\frac{ s_{1/6}}{s_{5/6}} \tilde{P}^+_3 \nonumber\\
{\cal FT}\left[P_{b_1} -\frac{m_{b_1} L}{2\pi}\cosh\theta\right]&=&
-\frac{ s_{1/2}}{s_{5/6}} (\tilde{P}^+_s +\tilde{P}^+_a)
-\frac{ s_{1/6}}{s_{5/6}} \tilde{P}^+_{b_1}
-\frac{2 c_{1/6} s_{1/2}}{s_{5/6}} \tilde{P}^+_{b_2}\nonumber\\
&&
+\frac{ s_{1/6}}{s_{5/6}} \tilde{P}^+_1
-\frac{ s_{2/3}}{s_{5/6}} \tilde{P}^+_{1_-}
+\frac{ 2 c_{1/3} s_{1/6}}{s_{5/6}} \tilde{P}^+_2
-\frac{ 2 c_{1/3} s_{1/6}}{s_{5/6}} \tilde{P}^+_3 \nonumber\\
{\cal FT}\left[P_{b_2} -\frac{m_{b_2} L}{2\pi}\cosh\theta\right]&=&
-\frac{4 c_{1/6}^2 s_{1/3}}{s_{5/6}} (\tilde{P}^+_s +\tilde{P}^+_a)
-\frac{2 c_{1/6} s_{1/2}}{s_{5/6}} \tilde{P}^+_{b_1}
+\frac{ c_{1/6}(s_{1/6}+s_{1/3}- 3 s_{5/6})}{c_{1/2} s_{5/6}} \tilde{P}^+_{b_2}
\nonumber\\
& & +\frac{ 2 c_{1/6} s_{1/2}}{s_{5/6}} \tilde{P}^+_1
-\frac{ 2 c_{1/6} s_{1/3}}{s_{5/6}} \tilde{P}^+_{1_-}
+\frac{ s_{1/3}}{s_{5/6}} \tilde{P}^+_2
-\frac{ s_{1/3}}{s_{5/6}} \tilde{P}^+_3 \nonumber\\
\tilde{P}_1 &=&
\frac{s_{1/2}}{s_{5/6}} (\tilde{P}^+_s +\tilde{P}^+_a)
+\frac{ s_{1/6}}{s_{5/6}} \tilde{P}^+_{b_1}
+\frac{2 c_{1/6} s_{1/2}}{s_{5/6}} \tilde{P}^+_{b_2}\nonumber\\
&&
-\frac{ s_{1/6}}{s_{5/6}} \tilde{P}^+_1
+\frac{ s_{2/3}}{s_{5/6}} \tilde{P}^+_{1_-}
-\frac{ 2 c_{1/3} s_{1/6}}{s_{5/6}} \tilde{P}^+_2
+\frac{ 2 c_{1/3} s_{1/6}}{s_{5/6}} \tilde{P}^+_3 \nonumber\\
\tilde{P}_{1_-} &=&
\frac{s_{1/3}}{s_{5/6}} (\tilde{P}^+_s +\tilde{P}^+_a)
+\frac{ 2 c_{1/3} s_{1/6}}{s_{5/6}} \tilde{P}^+_{b_1}
+\frac{2 c_{1/6} s_{1/3}}{s_{5/6}} \tilde{P}^+_{b_2}\nonumber\\
&&
-\frac{ s_{2/3}}{s_{5/6}} \tilde{P}^+_1
+\frac{ s_{1/6}}{s_{5/6}} \tilde{P}^+_{1_-}
+\frac{ 2 c_{1/2} s_{1/6}}{s_{5/6}} \tilde{P}^+_2
-\frac{ 2 c_{1/2} s_{1/6}}{s_{5/6}} \tilde{P}^+_3 \nonumber\\
\tilde{P}_2&=&
\frac{s_{1/6}}{s_{5/6}} (\tilde{P}^+_s +\tilde{P}^+_a)
+\frac{2 c_{1/3} s_{1/6}}{s_{5/6}} \tilde{P}^+_{b_1}
+\frac{s_{1/3}}{s_{5/6}} \tilde{P}^+_{b_2}
\nonumber\\
& &
-\frac{2 c_{1/3} s_{1/6}}{s_{5/6}} \tilde{P}^+_1
-\frac{2 c_{1/2} s_{1/6}}{s_{5/6}} \tilde{P}^+_{1_-}
+\frac{ c_{1/2} s_{1/6}}{c_{1/6} s_{5/6}} \tilde{P}^+_2
-\frac{ c_{1/2} s_{1/6}}{c_{1/6} s_{5/6}} \tilde{P}^+_3 \nonumber\\
\tilde{P}_3 &=& \tilde{P}_2
\end{eqnarray}
where $s_m = \sinh m\pi k$ and $c_m = \cosh m\pi k$ and the positive parity
of the strings $1_+$, $2_+$ and $3_+$ is understood in the notation.
Furthermore we never write the corresponding equation for the antisoliton
density being trivially $P_s = P_a$.
\end{widetext}
Simple manipulations with trigonometric identities results into a drastic
simplification of the system
\begin{widetext}
\begin{eqnarray}
{\cal FT}\left[P_s -\frac{m_s L}{2\pi}\cosh\theta\right] &=& 
-\frac{1}{2 c_{1/3}} (\tilde{P}^+_s +\tilde{P}_a^+) 
-\frac{1}{2 c_{1/3}} +\tilde{P}^+_{b_1} 
-\frac{c_{1/6}}{ c_{1/3}} +\tilde{P}^+_{b_2} 
-\frac{1}{2 c_{1/3}} +\tilde{P}^-_{1}
\nonumber\\
{\cal FT}\left[P_{b_2} -\frac{m_{b_2} L}{2\pi}\cosh\theta\right] &=& 
-\frac{c_{1/6}}{ c_{1/3}} (\tilde{P}^+_s +\tilde{P}_a^+) 
-\frac{c_{1/6}}{ c_{1/3}} +\tilde{P}^+_{b_1} 
-\frac{1}{ c_{1/3}} +\tilde{P}^+_{b_2} 
-\frac{c_{1/6}}{c_{1/3}} +\tilde{P}^-_{1}
\nonumber\\
{\cal FT}\left[P_{b_1} -\frac{m_{b_1} L}{2\pi}\cosh\theta\right] &=& 
-\frac{1}{2 c_{1/3}} (\tilde{P}^+_s +\tilde{P}_a^+) 
-\frac{1}{2 c_{1/3}} +\tilde{P}^+_{b_1} 
-\frac{c_{1/6}}{ c_{1/3}} +\tilde{P}^+_{b_2} 
-\frac{1}{2 c_{1/3}} +\tilde{P}^-_{1}
\nonumber\\
\tilde{P}_1 &=& 
\frac{1}{2 c_{1/3}} (\tilde{P}^+_s +\tilde{P}_a^+) 
+\frac{1}{2 c_{1/3}} +\tilde{P}^+_{b_1} 
+\frac{c_{1/6}}{ c_{1/3}} +\tilde{P}^+_{b_2} 
+\frac{1}{2 c_{1/3}} +\tilde{P}^-_{1}
-\frac{1}{2 c_{1/6}} +\tilde{P}^-_{1_-}
\nonumber\\
\tilde{P}_{1_-} &=& 
\frac{1}{2 c_{1/6}} \left[\tilde{P}^+_{b_1} +\tilde{P}^-_{1}
+\tilde{P}^+_{2}+\tilde{P}^+_{3}\right]
\nonumber\\
\tilde{P}_{2} &=& 
\frac{1}{2 c_{1/6}} \tilde{P}^-_{1_-}
\nonumber\\
\tilde{P}_3 &=& \tilde{P}_2\,\,\,\,.
\label{YYimproved}
\end{eqnarray}
\end{widetext}
The hole densities $P^-_i$ are defined as $P_i^- = P_i - P^+_i$.

We now have all the ingredients to make a thermodynamical analysis.
The internal energy is 
\begin{eqnarray}
U &=& \int d\theta \{m_s \cosh\theta [P^+_s(\theta) +P^+_a(\theta)]
\nonumber\\&&+ m_{b_1} \cosh\theta P^+_{b_1}(\theta)
+ m_{b_2} \cosh\theta P^+_{b_2}(\theta)\}
\end{eqnarray}
and the entropy, in the Stirling approximation,
\begin{equation}
S = \int d\theta \sum_i(P_i\log P_i + P^+_i \log P^+_i + P^-_i \log P^-_i)\,\,\,\,.
\end{equation}
Next, we minimizing the free energy $F = U - T S$ (here $T$ is the temperature) with respect to 
the densities given by
\begin{equation}
\rho^+_i=\{P^+_s,  P^+_a,  P^+_{b_i}, P^+_2,  P^+_3,  P^-_1, P^-_{1_-}\}.
\end{equation}
Taking into account the constraints of Eq. \eqref{YYimproved}, we end up to 
the following TBA system for the pseudoenergies $\epsilon_i(\theta)$ defined as
$\rho^+_i/P_i = e^{-\epsilon_i}/(1+e^{-\epsilon_i})$
\begin{widetext}
\begin{eqnarray}
\epsilon_s(\theta)&=& m_sR\cosh\theta + 
\hat\phi_1 \star \left[L_s + L_a + L_{b_1} -L_1\right] +\hat\phi_3 \star L_{b_2}
\nonumber\\
\epsilon_{b_2}(\theta)&=& m_{b_2}R\cosh\theta + 
\hat\phi_3 \star \left[L_s + L_a + L_{b_1} -L_1\right] +2\hat\phi_1 \star L_{b_2}
\nonumber\\
\epsilon_{b_1}(\theta)&=& m_{b_1}R\cosh\theta + 
\hat\phi_1 \star \left[L_s + L_a + L_{b_1} -L_1\right] +\hat\phi_3 \star L_{b_2}
-\phi_2 \star L_{1_-}
\nonumber\\
\epsilon_{1}(\theta)&=&  
\hat\phi_1 \star \left[L_s + L_a + L_{b_1} -L_1\right] +\hat\phi_3 \star L_{b_2}
-\hat\phi_2 \star L_{1_-}
\nonumber\\
\epsilon_{1_-}(\theta)&=&  
-\hat\phi_2 \star \left[L_{b_1} -L_1 +L_2 +L_3\right]
\nonumber\\
\epsilon_{2}(\theta)&=&  
-\hat\phi_2 \star L_{1_-}
\nonumber\\
\epsilon_{3}(\theta)&=&  
-\hat\phi_2 \star L_{1_-},
\end{eqnarray}
\end{widetext}
where $R$ is the inverse temperature. 
Here we have introduced the notation $L_i(\theta) = \log(1+ e^{-\epsilon_i})$
and the new kernels 
$\tilde{\phi}_1(k)=1/2\cosh\frac{\pi k}{3}$, 
$\tilde{\phi}_2(k)=1/2\cosh\frac{\pi k}{6}$ and 
$\tilde{\phi}_3(k)=\cosh\frac{\pi k}{6}/\cosh\frac{\pi k}{3}$. 

A further trivial analytic and algebraic manipulation gives the final form for 
the $\epsilon$-system
\begin{widetext}
\begin{eqnarray}
\epsilon_s(\theta)&=& 
\hat\varphi_2 \star {\cal L}_{b_2}
\nonumber\\
\epsilon_{b_2}(\theta)&=& 
\hat\varphi_2 \star\left[{\cal L}_s +{\cal L}_a +{\cal L}_{b_1} -{\cal L}_1\right] 
\nonumber\\
\epsilon_{b_1}(\theta)&=&
\hat\varphi_2 \star \left[{\cal L}_{b_2} -L_{2}\right] 
\nonumber\\
\epsilon_{1}(\theta)&=& - m_{b_1}R\cosh\theta 
+\hat\varphi_2 \star \left[{\cal L}_{b_2} -L_{1_-}\right]
\nonumber\\
\epsilon_{1_-}(\theta)&=&  
-\hat\varphi_2 \star \left[L_{b_1} -L_1 +L_2 +L_3\right]
\nonumber\\
\epsilon_{2}(\theta)&=&  
-\hat\varphi_2 \star L_{1_-}
\nonumber\\\label{tba_n_2}
\epsilon_{3}(\theta)&=&  
-\hat\varphi_2 \star L_{1_-}\,\,\,.
\end{eqnarray}
\end{widetext}
The functions ${\cal L}_i$ are the positive part of $\epsilon_i$, i.e.
${\cal L}_i = \log(1+ e^{\epsilon_i})$. The universal kernel is
$\tilde{\varphi}_2(k)=1/2\cosh\frac{\pi k}{6}$. The above procedure can be generalized to other integer values of $n$, leading to the equations presented in Eq. \eqref{TBA_eq}.

\bibliography{library_1}

\begin{thebibliography}{88}%
\makeatletter
\providecommand \@ifxundefined [1]{%
 \@ifx{#1\undefined}
}%
\providecommand \@ifnum [1]{%
 \ifnum #1\expandafter \@firstoftwo
 \else \expandafter \@secondoftwo
 \fi
}%
\providecommand \@ifx [1]{%
 \ifx #1\expandafter \@firstoftwo
 \else \expandafter \@secondoftwo
 \fi
}%
\providecommand \natexlab [1]{#1}%
\providecommand \enquote  [1]{``#1''}%
\providecommand \bibnamefont  [1]{#1}%
\providecommand \bibfnamefont [1]{#1}%
\providecommand \citenamefont [1]{#1}%
\providecommand \href@noop [0]{\@secondoftwo}%
\providecommand \href [0]{\begingroup \@sanitize@url \@href}%
\providecommand \@href[1]{\@@startlink{#1}\@@href}%
\providecommand \@@href[1]{\endgroup#1\@@endlink}%
\providecommand \@sanitize@url [0]{\catcode `\\12\catcode `\$12\catcode
  `\&12\catcode `\#12\catcode `\^12\catcode `\_12\catcode `\%12\relax}%
\providecommand \@@startlink[1]{}%
\providecommand \@@endlink[0]{}%
\providecommand \url  [0]{\begingroup\@sanitize@url \@url }%
\providecommand \@url [1]{\endgroup\@href {#1}{\urlprefix }}%
\providecommand \urlprefix  [0]{URL }%
\providecommand \Eprint [0]{\href }%
\providecommand \doibase [0]{http://dx.doi.org/}%
\providecommand \selectlanguage [0]{\@gobble}%
\providecommand \bibinfo  [0]{\@secondoftwo}%
\providecommand \bibfield  [0]{\@secondoftwo}%
\providecommand \translation [1]{[#1]}%
\providecommand \BibitemOpen [0]{}%
\providecommand \bibitemStop [0]{}%
\providecommand \bibitemNoStop [0]{.\EOS\space}%
\providecommand \EOS [0]{\spacefactor3000\relax}%
\providecommand \BibitemShut  [1]{\csname bibitem#1\endcsname}%
\let\auto@bib@innerbib\@empty
\bibitem [{\citenamefont {Feynman}(1982)}]{Feynman_1982}%
  \BibitemOpen
  \bibfield  {author} {\bibinfo {author} {\bibfnamefont {R.~P.}\ \bibnamefont
  {Feynman}},\ }\href@noop {} {\bibfield  {journal} {\bibinfo  {journal}
  {Inernational Journal of Theoretical Physics}\ }\textbf {\bibinfo {volume}
  {21}},\ \bibinfo {pages} {467} (\bibinfo {year} {1982})}\BibitemShut
  {NoStop}%
\bibitem [{\citenamefont {Lloyd}(1996)}]{Lloyd_1996}%
  \BibitemOpen
  \bibfield  {author} {\bibinfo {author} {\bibfnamefont {S.}~\bibnamefont
  {Lloyd}},\ }\href {\doibase 10.1126/science.279.5354.1113h} {\bibfield
  {journal} {\bibinfo  {journal} {Science}\ }\textbf {\bibinfo {volume}
  {273}},\ \bibinfo {pages} {1073} (\bibinfo {year} {1996})}\BibitemShut
  {NoStop}%
\bibitem [{\citenamefont {Abrams}\ and\ \citenamefont
  {Lloyd}(1997)}]{Lloyd1997}%
  \BibitemOpen
  \bibfield  {author} {\bibinfo {author} {\bibfnamefont {D.~S.}\ \bibnamefont
  {Abrams}}\ and\ \bibinfo {author} {\bibfnamefont {S.}~\bibnamefont {Lloyd}},\
  }\href {\doibase 10.1103/PhysRevLett.79.2586} {\bibfield  {journal} {\bibinfo
   {journal} {Phys. Rev. Lett.}\ }\textbf {\bibinfo {volume} {79}},\ \bibinfo
  {pages} {2586} (\bibinfo {year} {1997})}\BibitemShut {NoStop}%
\bibitem [{\citenamefont {Dou\ifmmode~\mbox{\c{c}}\else \c{c}\fi{}ot}\ \emph
  {et~al.}(2004)\citenamefont {Dou\ifmmode~\mbox{\c{c}}\else \c{c}\fi{}ot},
  \citenamefont {Ioffe},\ and\ \citenamefont {Vidal}}]{Doucot2004}%
  \BibitemOpen
  \bibfield  {author} {\bibinfo {author} {\bibfnamefont {B.}~\bibnamefont
  {Dou\ifmmode~\mbox{\c{c}}\else \c{c}\fi{}ot}}, \bibinfo {author}
  {\bibfnamefont {L.~B.}\ \bibnamefont {Ioffe}}, \ and\ \bibinfo {author}
  {\bibfnamefont {J.}~\bibnamefont {Vidal}},\ }\href {\doibase
  10.1103/PhysRevB.69.214501} {\bibfield  {journal} {\bibinfo  {journal} {Phys.
  Rev. B}\ }\textbf {\bibinfo {volume} {69}},\ \bibinfo {pages} {214501}
  (\bibinfo {year} {2004})}\BibitemShut {NoStop}%
\bibitem [{\citenamefont {B\"uchler}\ \emph {et~al.}(2005)\citenamefont
  {B\"uchler}, \citenamefont {Hermele}, \citenamefont {Huber}, \citenamefont
  {Fisher},\ and\ \citenamefont {Zoller}}]{Buchler2005}%
  \BibitemOpen
  \bibfield  {author} {\bibinfo {author} {\bibfnamefont {H.~P.}\ \bibnamefont
  {B\"uchler}}, \bibinfo {author} {\bibfnamefont {M.}~\bibnamefont {Hermele}},
  \bibinfo {author} {\bibfnamefont {S.~D.}\ \bibnamefont {Huber}}, \bibinfo
  {author} {\bibfnamefont {M.~P.~A.}\ \bibnamefont {Fisher}}, \ and\ \bibinfo
  {author} {\bibfnamefont {P.}~\bibnamefont {Zoller}},\ }\href {\doibase
  10.1103/PhysRevLett.95.040402} {\bibfield  {journal} {\bibinfo  {journal}
  {Phys. Rev. Lett.}\ }\textbf {\bibinfo {volume} {95}},\ \bibinfo {pages}
  {040402} (\bibinfo {year} {2005})}\BibitemShut {NoStop}%
\bibitem [{\citenamefont {Cirac}\ \emph {et~al.}(2010)\citenamefont {Cirac},
  \citenamefont {Maraner},\ and\ \citenamefont {Pachos}}]{Cirac2010}%
  \BibitemOpen
  \bibfield  {author} {\bibinfo {author} {\bibfnamefont {J.~I.}\ \bibnamefont
  {Cirac}}, \bibinfo {author} {\bibfnamefont {P.}~\bibnamefont {Maraner}}, \
  and\ \bibinfo {author} {\bibfnamefont {J.~K.}\ \bibnamefont {Pachos}},\
  }\href {\doibase 10.1103/PhysRevLett.105.190403} {\bibfield  {journal}
  {\bibinfo  {journal} {Phys. Rev. Lett.}\ }\textbf {\bibinfo {volume} {105}},\
  \bibinfo {pages} {190403} (\bibinfo {year} {2010})}\BibitemShut {NoStop}%
\bibitem [{\citenamefont {Casanova}\ \emph {et~al.}(2011)\citenamefont
  {Casanova}, \citenamefont {Lamata}, \citenamefont {Egusquiza}, \citenamefont
  {Gerritsma}, \citenamefont {Roos}, \citenamefont {Garc\'{\i}a-Ripoll},\ and\
  \citenamefont {Solano}}]{Casanova2011}%
  \BibitemOpen
  \bibfield  {author} {\bibinfo {author} {\bibfnamefont {J.}~\bibnamefont
  {Casanova}}, \bibinfo {author} {\bibfnamefont {L.}~\bibnamefont {Lamata}},
  \bibinfo {author} {\bibfnamefont {I.~L.}\ \bibnamefont {Egusquiza}}, \bibinfo
  {author} {\bibfnamefont {R.}~\bibnamefont {Gerritsma}}, \bibinfo {author}
  {\bibfnamefont {C.~F.}\ \bibnamefont {Roos}}, \bibinfo {author}
  {\bibfnamefont {J.~J.}\ \bibnamefont {Garc\'{\i}a-Ripoll}}, \ and\ \bibinfo
  {author} {\bibfnamefont {E.}~\bibnamefont {Solano}},\ }\href {\doibase
  10.1103/PhysRevLett.107.260501} {\bibfield  {journal} {\bibinfo  {journal}
  {Phys. Rev. Lett.}\ }\textbf {\bibinfo {volume} {107}},\ \bibinfo {pages}
  {260501} (\bibinfo {year} {2011})}\BibitemShut {NoStop}%
\bibitem [{\citenamefont {Jordan}\ \emph {et~al.}(2014)\citenamefont {Jordan},
  \citenamefont {Lee},\ and\ \citenamefont {Preskill}}]{Jordan2011}%
  \BibitemOpen
  \bibfield  {author} {\bibinfo {author} {\bibfnamefont {S.~P.}\ \bibnamefont
  {Jordan}}, \bibinfo {author} {\bibfnamefont {K.~S.~M.}\ \bibnamefont {Lee}},
  \ and\ \bibinfo {author} {\bibfnamefont {J.}~\bibnamefont {Preskill}},\
  }\href@noop {} {\bibfield  {journal} {\bibinfo  {journal} {Quant. Inf.
  Comput.}\ }\textbf {\bibinfo {volume} {14}} (\bibinfo {year}
  {2014})}\BibitemShut {NoStop}%
\bibitem [{\citenamefont {Jordan}\ \emph {et~al.}(2012)\citenamefont {Jordan},
  \citenamefont {Lee},\ and\ \citenamefont {Preskill}}]{Jordan2012}%
  \BibitemOpen
  \bibfield  {author} {\bibinfo {author} {\bibfnamefont {S.~P.}\ \bibnamefont
  {Jordan}}, \bibinfo {author} {\bibfnamefont {K.~S.~M.}\ \bibnamefont {Lee}},
  \ and\ \bibinfo {author} {\bibfnamefont {J.}~\bibnamefont {Preskill}},\
  }\href {\doibase 10.1126/science.1217069} {\bibfield  {journal} {\bibinfo
  {journal} {Science}\ }\textbf {\bibinfo {volume} {336}},\ \bibinfo {pages}
  {1130} (\bibinfo {year} {2012})},\ \Eprint
  {http://arxiv.org/abs/http://science.sciencemag.org/content/336/6085/1130.full.pdf}
  {http://science.sciencemag.org/content/336/6085/1130.full.pdf} \BibitemShut
  {NoStop}%
\bibitem [{\citenamefont {Houck}\ \emph {et~al.}(2012)\citenamefont {Houck},
  \citenamefont {T{\"u}reci},\ and\ \citenamefont {Koch}}]{Houck2012}%
  \BibitemOpen
  \bibfield  {author} {\bibinfo {author} {\bibfnamefont {A.~A.}\ \bibnamefont
  {Houck}}, \bibinfo {author} {\bibfnamefont {H.~E.}\ \bibnamefont
  {T{\"u}reci}}, \ and\ \bibinfo {author} {\bibfnamefont {J.}~\bibnamefont
  {Koch}},\ }\href {https://doi.org/10.1038/nphys2251} {\bibfield  {journal}
  {\bibinfo  {journal} {Nature Physics}\ }\textbf {\bibinfo {volume} {8}},\
  \bibinfo {pages} {292 EP } (\bibinfo {year} {2012})}\BibitemShut {NoStop}%
\bibitem [{\citenamefont {Devoret}\ and\ \citenamefont
  {Schoelkopf}(2013)}]{Devoret_Schoelkopf_2013}%
  \BibitemOpen
  \bibfield  {author} {\bibinfo {author} {\bibfnamefont {M.~H.}\ \bibnamefont
  {Devoret}}\ and\ \bibinfo {author} {\bibfnamefont {R.~J.}\ \bibnamefont
  {Schoelkopf}},\ }\href {\doibase 10.1126/science.1231930} {\bibfield
  {journal} {\bibinfo  {journal} {Science}\ }\textbf {\bibinfo {volume}
  {339}},\ \bibinfo {pages} {1169} (\bibinfo {year} {2013})}\BibitemShut
  {NoStop}%
\bibitem [{\citenamefont {Fendley}\ \emph
  {et~al.}(1995{\natexlab{a}})\citenamefont {Fendley}, \citenamefont {Ludwig},\
  and\ \citenamefont {Saleur}}]{Fendley1995}%
  \BibitemOpen
  \bibfield  {author} {\bibinfo {author} {\bibfnamefont {P.}~\bibnamefont
  {Fendley}}, \bibinfo {author} {\bibfnamefont {A.~W.~W.}\ \bibnamefont
  {Ludwig}}, \ and\ \bibinfo {author} {\bibfnamefont {H.}~\bibnamefont
  {Saleur}},\ }\href {\doibase 10.1103/PhysRevB.52.8934} {\bibfield  {journal}
  {\bibinfo  {journal} {Phys. Rev. B}\ }\textbf {\bibinfo {volume} {52}},\
  \bibinfo {pages} {8934} (\bibinfo {year} {1995}{\natexlab{a}})}\BibitemShut
  {NoStop}%
\bibitem [{\citenamefont {Fendley}\ \emph
  {et~al.}(1995{\natexlab{b}})\citenamefont {Fendley}, \citenamefont {Ludwig},\
  and\ \citenamefont {Saleur}}]{Fendley1995a}%
  \BibitemOpen
  \bibfield  {author} {\bibinfo {author} {\bibfnamefont {P.}~\bibnamefont
  {Fendley}}, \bibinfo {author} {\bibfnamefont {A.~W.~W.}\ \bibnamefont
  {Ludwig}}, \ and\ \bibinfo {author} {\bibfnamefont {H.}~\bibnamefont
  {Saleur}},\ }\href {\doibase 10.1103/PhysRevLett.74.3005} {\bibfield
  {journal} {\bibinfo  {journal} {Phys. Rev. Lett.}\ }\textbf {\bibinfo
  {volume} {74}},\ \bibinfo {pages} {3005} (\bibinfo {year}
  {1995}{\natexlab{b}})}\BibitemShut {NoStop}%
\bibitem [{\citenamefont {Milliken}\ \emph {et~al.}(1996)\citenamefont
  {Milliken}, \citenamefont {Umbach},\ and\ \citenamefont
  {Webb}}]{Milliken1996}%
  \BibitemOpen
  \bibfield  {author} {\bibinfo {author} {\bibfnamefont {F.}~\bibnamefont
  {Milliken}}, \bibinfo {author} {\bibfnamefont {C.}~\bibnamefont {Umbach}}, \
  and\ \bibinfo {author} {\bibfnamefont {R.}~\bibnamefont {Webb}},\ }\href
  {\doibase https://doi.org/10.1016/0038-1098(95)00181-6} {\bibfield  {journal}
  {\bibinfo  {journal} {Solid State Communications}\ }\textbf {\bibinfo
  {volume} {97}},\ \bibinfo {pages} {309 } (\bibinfo {year}
  {1996})}\BibitemShut {NoStop}%
\bibitem [{\citenamefont {Lesage}\ \emph {et~al.}(1997)\citenamefont {Lesage},
  \citenamefont {Saleur},\ and\ \citenamefont {Simonetti}}]{Lesage1997}%
  \BibitemOpen
  \bibfield  {author} {\bibinfo {author} {\bibfnamefont {F.}~\bibnamefont
  {Lesage}}, \bibinfo {author} {\bibfnamefont {H.}~\bibnamefont {Saleur}}, \
  and\ \bibinfo {author} {\bibfnamefont {P.}~\bibnamefont {Simonetti}},\ }\href
  {\doibase 10.1103/PhysRevB.56.7598} {\bibfield  {journal} {\bibinfo
  {journal} {Phys. Rev. B}\ }\textbf {\bibinfo {volume} {56}},\ \bibinfo
  {pages} {7598} (\bibinfo {year} {1997})}\BibitemShut {NoStop}%
\bibitem [{\citenamefont {Giuliano}\ and\ \citenamefont
  {Sodano}(2005)}]{Giulliano2005}%
  \BibitemOpen
  \bibfield  {author} {\bibinfo {author} {\bibfnamefont {D.}~\bibnamefont
  {Giuliano}}\ and\ \bibinfo {author} {\bibfnamefont {P.}~\bibnamefont
  {Sodano}},\ }\href {\doibase https://doi.org/10.1016/j.nuclphysb.2005.01.037}
  {\bibfield  {journal} {\bibinfo  {journal} {Nuclear Physics B}\ }\textbf
  {\bibinfo {volume} {711}},\ \bibinfo {pages} {480 } (\bibinfo {year}
  {2005})}\BibitemShut {NoStop}%
\bibitem [{\citenamefont {Gritsev}\ \emph {et~al.}(2007)\citenamefont
  {Gritsev}, \citenamefont {Polkovnikov},\ and\ \citenamefont
  {Demler}}]{Gritsev2007}%
  \BibitemOpen
  \bibfield  {author} {\bibinfo {author} {\bibfnamefont {V.}~\bibnamefont
  {Gritsev}}, \bibinfo {author} {\bibfnamefont {A.}~\bibnamefont
  {Polkovnikov}}, \ and\ \bibinfo {author} {\bibfnamefont {E.}~\bibnamefont
  {Demler}},\ }\href {\doibase 10.1103/PhysRevB.75.174511} {\bibfield
  {journal} {\bibinfo  {journal} {Phys. Rev. B}\ }\textbf {\bibinfo {volume}
  {75}},\ \bibinfo {pages} {174511} (\bibinfo {year} {2007})}\BibitemShut
  {NoStop}%
\bibitem [{\citenamefont {Bukhvostov}\ and\ \citenamefont
  {Lipatov}(1981)}]{Bukhvostov1980}%
  \BibitemOpen
  \bibfield  {author} {\bibinfo {author} {\bibfnamefont {A.~P.}\ \bibnamefont
  {Bukhvostov}}\ and\ \bibinfo {author} {\bibfnamefont {L.~N.}\ \bibnamefont
  {Lipatov}},\ }\href {\doibase 10.1016/0550-3213(81)90157-7} {\bibfield
  {journal} {\bibinfo  {journal} {Nucl. Phys.}\ }\textbf {\bibinfo {volume}
  {B180}},\ \bibinfo {pages} {116} (\bibinfo {year} {1981})},\ \bibinfo {note}
  {[Pisma Zh. Eksp. Teor. Fiz.31,138(1980)]}\BibitemShut {NoStop}%
\bibitem [{\citenamefont {Saleur}(1999)}]{Saleur1999}%
  \BibitemOpen
  \bibfield  {author} {\bibinfo {author} {\bibfnamefont {H.}~\bibnamefont
  {Saleur}},\ }\href {\doibase 10.1088/0305-4470/32/18/102} {\bibfield
  {journal} {\bibinfo  {journal} {Journal of Physics A: Mathematical and
  General}\ }\textbf {\bibinfo {volume} {32}},\ \bibinfo {pages} {L207}
  (\bibinfo {year} {1999})}\BibitemShut {NoStop}%
\bibitem [{\citenamefont {Giamarchi}(2003)}]{Giamarchi2003}%
  \BibitemOpen
  \bibfield  {author} {\bibinfo {author} {\bibfnamefont {T.}~\bibnamefont
  {Giamarchi}},\ }\href {https://books.google.de/books?id=GVeuKZLGMZ0C} {\emph
  {\bibinfo {title} {Quantum Physics in One Dimension}}},\ International Series
  of Monographs on Physics\ (\bibinfo  {publisher} {Clarendon Press},\ \bibinfo
  {year} {2003})\BibitemShut {NoStop}%
\bibitem [{\citenamefont {Gogolin}\ \emph {et~al.}(2004)\citenamefont
  {Gogolin}, \citenamefont {Nersesyan},\ and\ \citenamefont
  {Tsvelik}}]{Gogolin2004}%
  \BibitemOpen
  \bibfield  {author} {\bibinfo {author} {\bibfnamefont {A.}~\bibnamefont
  {Gogolin}}, \bibinfo {author} {\bibfnamefont {A.}~\bibnamefont {Nersesyan}},
  \ and\ \bibinfo {author} {\bibfnamefont {A.}~\bibnamefont {Tsvelik}},\ }\href
  {https://books.google.de/books?id=BZDfFIpCoaAC} {\emph {\bibinfo {title}
  {Bosonization and Strongly Correlated Systems}}}\ (\bibinfo  {publisher}
  {Cambridge University Press},\ \bibinfo {year} {2004})\BibitemShut {NoStop}%
\bibitem [{\citenamefont {Fisher}\ \emph {et~al.}(1989)\citenamefont {Fisher},
  \citenamefont {Weichman}, \citenamefont {Grinstein},\ and\ \citenamefont
  {Fisher}}]{Fisher1989}%
  \BibitemOpen
  \bibfield  {author} {\bibinfo {author} {\bibfnamefont {M.~P.~A.}\
  \bibnamefont {Fisher}}, \bibinfo {author} {\bibfnamefont {P.~B.}\
  \bibnamefont {Weichman}}, \bibinfo {author} {\bibfnamefont {G.}~\bibnamefont
  {Grinstein}}, \ and\ \bibinfo {author} {\bibfnamefont {D.~S.}\ \bibnamefont
  {Fisher}},\ }\href {\doibase 10.1103/PhysRevB.40.546} {\bibfield  {journal}
  {\bibinfo  {journal} {Phys. Rev. B}\ }\textbf {\bibinfo {volume} {40}},\
  \bibinfo {pages} {546} (\bibinfo {year} {1989})}\BibitemShut {NoStop}%
\bibitem [{\citenamefont {Sachdev}(2011)}]{Sachdev2011}%
  \BibitemOpen
  \bibfield  {author} {\bibinfo {author} {\bibfnamefont {S.}~\bibnamefont
  {Sachdev}},\ }\href {https://books.google.de/books?id=F3IkpxwpqSgC} {\emph
  {\bibinfo {title} {{Quantum Phase Transitions}}}}\ (\bibinfo  {publisher}
  {Cambridge University Press},\ \bibinfo {year} {2011})\BibitemShut {NoStop}%
\bibitem [{\citenamefont {Tinkham}(2004)}]{Tinkham2004}%
  \BibitemOpen
  \bibfield  {author} {\bibinfo {author} {\bibfnamefont {M.}~\bibnamefont
  {Tinkham}},\ }\href {https://books.google.de/books?id=k6AO9nRYbioC} {\emph
  {\bibinfo {title} {Introduction to Superconductivity: Second Edition}}},\
  Dover Books on Physics\ (\bibinfo  {publisher} {Dover Publications},\
  \bibinfo {year} {2004})\BibitemShut {NoStop}%
\bibitem [{\citenamefont {Devoret}(1997)}]{Devoret_1997}%
  \BibitemOpen
  \bibfield  {author} {\bibinfo {author} {\bibfnamefont {M.~H.}\ \bibnamefont
  {Devoret}},\ }\href@noop {} {\emph {\bibinfo {title} {{Quantum Fluctuations
  in Electrical Circuits (Les Houches Session LXIII)}}}}\ (\bibinfo
  {publisher} {Elsevier},\ \bibinfo {year} {1997})\ pp.\ \bibinfo {pages}
  {351--386}\BibitemShut {NoStop}%
\bibitem [{Note1()}]{Note1}%
  \BibitemOpen
  \bibinfo {note} {This is true also if one tries to ``backtrack'' to the
  fermionic formulation, where ``non-Fermi liquid'' fixed points are then
  encountered}\BibitemShut {NoStop}%
\bibitem [{\citenamefont {Iagolnitzer}(1978)}]{Iagolnitzer1978}%
  \BibitemOpen
  \bibfield  {author} {\bibinfo {author} {\bibfnamefont {D.}~\bibnamefont
  {Iagolnitzer}},\ }\href {\doibase 10.1103/PhysRevD.18.1275} {\bibfield
  {journal} {\bibinfo  {journal} {Phys. Rev. D}\ }\textbf {\bibinfo {volume}
  {18}},\ \bibinfo {pages} {1275} (\bibinfo {year} {1978})}\BibitemShut
  {NoStop}%
\bibitem [{\citenamefont {Zamolodchikov}\ and\ \citenamefont
  {Zamolodchikov}(1979)}]{Zamolodchikov1979}%
  \BibitemOpen
  \bibfield  {author} {\bibinfo {author} {\bibfnamefont {A.~B.}\ \bibnamefont
  {Zamolodchikov}}\ and\ \bibinfo {author} {\bibfnamefont {A.~B.}\ \bibnamefont
  {Zamolodchikov}},\ }\href {\doibase
  https://doi.org/10.1016/0003-4916(79)90391-9} {\bibfield  {journal} {\bibinfo
   {journal} {Annals of Physics}\ }\textbf {\bibinfo {volume} {120}},\ \bibinfo
  {pages} {253 } (\bibinfo {year} {1979})}\BibitemShut {NoStop}%
\bibitem [{\citenamefont {Zamolodchikov}(1990)}]{Zamolodchikov1990a}%
  \BibitemOpen
  \bibfield  {author} {\bibinfo {author} {\bibfnamefont {A.}~\bibnamefont
  {Zamolodchikov}},\ }\href {\doibase
  https://doi.org/10.1016/0550-3213(90)90333-9} {\bibfield  {journal} {\bibinfo
   {journal} {Nuclear Physics B}\ }\textbf {\bibinfo {volume} {342}},\ \bibinfo
  {pages} {695 } (\bibinfo {year} {1990})}\BibitemShut {NoStop}%
\bibitem [{\citenamefont {Smirnov}(1992)}]{Smirnov1992}%
  \BibitemOpen
  \bibfield  {author} {\bibinfo {author} {\bibfnamefont {F.}~\bibnamefont
  {Smirnov}},\ }\href {https://books.google.de/books?id=pwMQkdBZ7YMC} {\emph
  {\bibinfo {title} {Form Factors in Completely Integrable Models of Quantum
  Field Theory}}},\ Advanced series in mathematical physics\ (\bibinfo
  {publisher} {World Scientific},\ \bibinfo {year} {1992})\BibitemShut
  {NoStop}%
\bibitem [{\citenamefont {Senthil}\ and\ \citenamefont
  {Fisher}(2000)}]{Senthil2000}%
  \BibitemOpen
  \bibfield  {author} {\bibinfo {author} {\bibfnamefont {T.}~\bibnamefont
  {Senthil}}\ and\ \bibinfo {author} {\bibfnamefont {M.~P.~A.}\ \bibnamefont
  {Fisher}},\ }\href {\doibase 10.1103/PhysRevB.62.7850} {\bibfield  {journal}
  {\bibinfo  {journal} {Phys. Rev. B}\ }\textbf {\bibinfo {volume} {62}},\
  \bibinfo {pages} {7850} (\bibinfo {year} {2000})}\BibitemShut {NoStop}%
\bibitem [{\citenamefont {Xu}\ and\ \citenamefont {Fu}(2010)}]{Xu2010}%
  \BibitemOpen
  \bibfield  {author} {\bibinfo {author} {\bibfnamefont {C.}~\bibnamefont
  {Xu}}\ and\ \bibinfo {author} {\bibfnamefont {L.}~\bibnamefont {Fu}},\ }\href
  {\doibase 10.1103/PhysRevB.81.134435} {\bibfield  {journal} {\bibinfo
  {journal} {Phys. Rev. B}\ }\textbf {\bibinfo {volume} {81}},\ \bibinfo
  {pages} {1} (\bibinfo {year} {2010})}\BibitemShut {NoStop}%
\bibitem [{\citenamefont {Roy}\ \emph {et~al.}(2017)\citenamefont {Roy},
  \citenamefont {Terhal},\ and\ \citenamefont {Hassler}}]{Roy2017}%
  \BibitemOpen
  \bibfield  {author} {\bibinfo {author} {\bibfnamefont {A.}~\bibnamefont
  {Roy}}, \bibinfo {author} {\bibfnamefont {B.~M.}\ \bibnamefont {Terhal}}, \
  and\ \bibinfo {author} {\bibfnamefont {F.}~\bibnamefont {Hassler}},\ }\href
  {\doibase 10.1103/PhysRevLett.119.180508} {\bibfield  {journal} {\bibinfo
  {journal} {Phys. Rev. Lett.}\ }\textbf {\bibinfo {volume} {119}},\ \bibinfo
  {pages} {180508} (\bibinfo {year} {2017})}\BibitemShut {NoStop}%
\bibitem [{\citenamefont {Nazarov}\ and\ \citenamefont
  {Blanter}(2009)}]{Nazarov_Blanter_2009}%
  \BibitemOpen
  \bibfield  {author} {\bibinfo {author} {\bibfnamefont {Y.~V.}\ \bibnamefont
  {Nazarov}}\ and\ \bibinfo {author} {\bibfnamefont {Y.~M.}\ \bibnamefont
  {Blanter}},\ }\href {\doibase 10.1017/CBO9780511626906} {\emph {\bibinfo
  {title} {Quantum Transport: Introduction to Nanoscience}}}\ (\bibinfo
  {publisher} {Cambridge University Press},\ \bibinfo {year}
  {2009})\BibitemShut {NoStop}%
\bibitem [{\citenamefont {Rajaraman}(1982)}]{Rajaraman1982}%
  \BibitemOpen
  \bibfield  {author} {\bibinfo {author} {\bibfnamefont {R.}~\bibnamefont
  {Rajaraman}},\ }\href {https://books.google.de/books?id=1XucQgAACAAJ} {\emph
  {\bibinfo {title} {Solitons and Instantons: An Introduction to Solitons and
  Instantons in Quantum Field Theory}}},\ North-Holland personal library\
  (\bibinfo  {publisher} {North-Holland Publishing Company},\ \bibinfo {year}
  {1982})\BibitemShut {NoStop}%
\bibitem [{\citenamefont {Ablowitz}\ and\ \citenamefont
  {Segur}(2006)}]{Ablowitz2006}%
  \BibitemOpen
  \bibfield  {author} {\bibinfo {author} {\bibfnamefont {M.}~\bibnamefont
  {Ablowitz}}\ and\ \bibinfo {author} {\bibfnamefont {H.}~\bibnamefont
  {Segur}},\ }\href {https://books.google.de/books?id=Bzu4XAUpFZUC} {\emph
  {\bibinfo {title} {Solitons and the Inverse Scattering Transform}}},\ SIAM
  Studies in Applied Mathematics\ (\bibinfo  {publisher} {Society for
  Industrial and Applied Mathematics},\ \bibinfo {year} {2006})\BibitemShut
  {NoStop}%
\bibitem [{\citenamefont {Mandelstam}(1975)}]{Mandelstam1975}%
  \BibitemOpen
  \bibfield  {author} {\bibinfo {author} {\bibfnamefont {S.}~\bibnamefont
  {Mandelstam}},\ }\href {\doibase 10.1103/PhysRevD.11.3026} {\bibfield
  {journal} {\bibinfo  {journal} {Phys. Rev. D}\ }\textbf {\bibinfo {volume}
  {11}},\ \bibinfo {pages} {3026} (\bibinfo {year} {1975})}\BibitemShut
  {NoStop}%
\bibitem [{\citenamefont {Ghoshal}\ and\ \citenamefont
  {Zamolodchikov}(1994)}]{Ghoshal1994}%
  \BibitemOpen
  \bibfield  {author} {\bibinfo {author} {\bibfnamefont {S.}~\bibnamefont
  {Ghoshal}}\ and\ \bibinfo {author} {\bibfnamefont {A.}~\bibnamefont
  {Zamolodchikov}},\ }\href {\doibase 10.1142/S0217751X94001552} {\bibfield
  {journal} {\bibinfo  {journal} {International Journal of Modern Physics A}\
  }\textbf {\bibinfo {volume} {09}},\ \bibinfo {pages} {3841} (\bibinfo {year}
  {1994})},\ \Eprint
  {http://arxiv.org/abs/https://doi.org/10.1142/S0217751X94001552}
  {https://doi.org/10.1142/S0217751X94001552} \BibitemShut {NoStop}%
\bibitem [{\citenamefont {Zarembo}(2017)}]{Zarembo2017}%
  \BibitemOpen
  \bibfield  {author} {\bibinfo {author} {\bibfnamefont {K.}~\bibnamefont
  {Zarembo}},\ }in\ \href@noop {} {\emph {\bibinfo {booktitle} {{Les Houches
  Summer School: Integrability: From Statistical Systems to Gauge Theory Les
  Houches, France, June 6-July 1, 2016}}}}\ (\bibinfo {year} {2017})\ \Eprint
  {http://arxiv.org/abs/1712.07725} {arXiv:1712.07725 [hep-th]} \BibitemShut
  {NoStop}%
\bibitem [{\citenamefont {Bernard}\ and\ \citenamefont
  {LeClair}(1991)}]{Bernard1991}%
  \BibitemOpen
  \bibfield  {author} {\bibinfo {author} {\bibfnamefont {D.}~\bibnamefont
  {Bernard}}\ and\ \bibinfo {author} {\bibfnamefont {A.}~\bibnamefont
  {LeClair}},\ }\href {https://projecteuclid.org:443/euclid.cmp/1104248491}
  {\bibfield  {journal} {\bibinfo  {journal} {Comm. Math. Phys.}\ }\textbf
  {\bibinfo {volume} {142}},\ \bibinfo {pages} {99} (\bibinfo {year}
  {1991})}\BibitemShut {NoStop}%
\bibitem [{\citenamefont {Belavin}\ \emph {et~al.}(1984)\citenamefont
  {Belavin}, \citenamefont {Polyakov},\ and\ \citenamefont
  {Zamolodchikov}}]{Belavin1984}%
  \BibitemOpen
  \bibfield  {author} {\bibinfo {author} {\bibfnamefont {A.}~\bibnamefont
  {Belavin}}, \bibinfo {author} {\bibfnamefont {A.}~\bibnamefont {Polyakov}}, \
  and\ \bibinfo {author} {\bibfnamefont {A.}~\bibnamefont {Zamolodchikov}},\
  }\href {\doibase https://doi.org/10.1016/0550-3213(84)90052-X} {\bibfield
  {journal} {\bibinfo  {journal} {Nuclear Physics B}\ }\textbf {\bibinfo
  {volume} {241}},\ \bibinfo {pages} {333 } (\bibinfo {year}
  {1984})}\BibitemShut {NoStop}%
\bibitem [{\citenamefont {Zamolodchikov}(1987)}]{Zamolodchikov1987}%
  \BibitemOpen
  \bibfield  {author} {\bibinfo {author} {\bibfnamefont {A.~B.}\ \bibnamefont
  {Zamolodchikov}},\ }\href@noop {} {\bibfield  {journal} {\bibinfo  {journal}
  {JETP Lett.}\ }\textbf {\bibinfo {volume} {46}},\ \bibinfo {pages} {160}
  (\bibinfo {year} {1987})},\ \bibinfo {note} {[Pisma Zh. Eksp. Teor.
  Fiz.46,129(1987)]}\BibitemShut {NoStop}%
\bibitem [{\citenamefont {Zamolodchikov}(1989)}]{Zamolodchikov1989}%
  \BibitemOpen
  \bibfield  {author} {\bibinfo {author} {\bibfnamefont {A.}~\bibnamefont
  {Zamolodchikov}},\ }in\ \href {\doibase
  https://doi.org/10.1016/B978-0-12-385342-4.50022-6} {\emph {\bibinfo
  {booktitle} {Integrable Sys Quantum Field Theory}}},\ \bibinfo {editor}
  {edited by\ \bibinfo {editor} {\bibfnamefont {M.}~\bibnamefont {Jimbo}},
  \bibinfo {editor} {\bibfnamefont {T.}~\bibnamefont {Miwa}}, \ and\ \bibinfo
  {editor} {\bibfnamefont {A.}~\bibnamefont {Tsuchiya}}}\ (\bibinfo
  {publisher} {Academic Press},\ \bibinfo {address} {San Diego},\ \bibinfo
  {year} {1989})\ pp.\ \bibinfo {pages} {641 -- 674}\BibitemShut {NoStop}%
\bibitem [{\citenamefont {Francesco}\ \emph {et~al.}(1997)\citenamefont
  {Francesco}, \citenamefont {Di~Francesco}, \citenamefont {Mathieu},
  \citenamefont {S{\'e}n{\'e}chal},\ and\ \citenamefont
  {Senechal}}]{diFrancesco1997}%
  \BibitemOpen
  \bibfield  {author} {\bibinfo {author} {\bibfnamefont {P.}~\bibnamefont
  {Francesco}}, \bibinfo {author} {\bibfnamefont {P.}~\bibnamefont
  {Di~Francesco}}, \bibinfo {author} {\bibfnamefont {P.}~\bibnamefont
  {Mathieu}}, \bibinfo {author} {\bibfnamefont {D.}~\bibnamefont
  {S{\'e}n{\'e}chal}}, \ and\ \bibinfo {author} {\bibfnamefont
  {D.}~\bibnamefont {Senechal}},\ }\href
  {https://books.google.de/books?id=keUrdME5rhIC} {\emph {\bibinfo {title}
  {Conformal Field Theory}}},\ Graduate Texts in Contemporary Physics\
  (\bibinfo  {publisher} {Springer},\ \bibinfo {year} {1997})\BibitemShut
  {NoStop}%
\bibitem [{\citenamefont {Mussardo}(2010)}]{Mussardo2010}%
  \BibitemOpen
  \bibfield  {author} {\bibinfo {author} {\bibfnamefont {G.}~\bibnamefont
  {Mussardo}},\ }\href {https://books.google.de/books?id=fakVDAAAQBAJ} {\emph
  {\bibinfo {title} {Statistical Field Theory: An Introduction to Exactly
  Solved Models in Statistical Physics}}},\ Oxford Graduate Texts\ (\bibinfo
  {publisher} {OUP Oxford},\ \bibinfo {year} {2010})\BibitemShut {NoStop}%
\bibitem [{\citenamefont {Affleck}\ \emph {et~al.}(2001)\citenamefont
  {Affleck}, \citenamefont {Oshikawa},\ and\ \citenamefont
  {Saleur}}]{Affleck2001}%
  \BibitemOpen
  \bibfield  {author} {\bibinfo {author} {\bibfnamefont {I.}~\bibnamefont
  {Affleck}}, \bibinfo {author} {\bibfnamefont {M.}~\bibnamefont {Oshikawa}}, \
  and\ \bibinfo {author} {\bibfnamefont {H.}~\bibnamefont {Saleur}},\ }\href
  {\doibase https://doi.org/10.1016/S0550-3213(00)00499-5} {\bibfield
  {journal} {\bibinfo  {journal} {Nuclear Physics B}\ }\textbf {\bibinfo
  {volume} {594}},\ \bibinfo {pages} {535 } (\bibinfo {year}
  {2001})}\BibitemShut {NoStop}%
\bibitem [{\citenamefont {Ameduri}\ and\ \citenamefont
  {Efthimiou}(1998)}]{Ameduri1998}%
  \BibitemOpen
  \bibfield  {author} {\bibinfo {author} {\bibfnamefont {M.}~\bibnamefont
  {Ameduri}}\ and\ \bibinfo {author} {\bibfnamefont {C.~J.}\ \bibnamefont
  {Efthimiou}},\ }\href {\doibase 10.2991/jnmp.1998.5.2.4} {\bibfield
  {journal} {\bibinfo  {journal} {Journal of Nonlinear Mathematical Physics}\
  }\textbf {\bibinfo {volume} {5}},\ \bibinfo {pages} {132} (\bibinfo {year}
  {1998})},\ \Eprint
  {http://arxiv.org/abs/https://doi.org/10.2991/jnmp.1998.5.2.4}
  {https://doi.org/10.2991/jnmp.1998.5.2.4} \BibitemShut {NoStop}%
\bibitem [{\citenamefont {Fateev}(1996)}]{Fateev1996}%
  \BibitemOpen
  \bibfield  {author} {\bibinfo {author} {\bibfnamefont {V.~A.}\ \bibnamefont
  {Fateev}},\ }\href {\doibase 10.1016/0550-3213(96)00256-8} {\bibfield
  {journal} {\bibinfo  {journal} {Nucl. Phys.}\ }\textbf {\bibinfo {volume}
  {B473}},\ \bibinfo {pages} {509} (\bibinfo {year} {1996})}\BibitemShut
  {NoStop}%
\bibitem [{\citenamefont {Lesage}\ \emph {et~al.}(1998)\citenamefont {Lesage},
  \citenamefont {Saleur},\ and\ \citenamefont {Simonetti}}]{Lesage1998}%
  \BibitemOpen
  \bibfield  {author} {\bibinfo {author} {\bibfnamefont {F.}~\bibnamefont
  {Lesage}}, \bibinfo {author} {\bibfnamefont {H.}~\bibnamefont {Saleur}}, \
  and\ \bibinfo {author} {\bibfnamefont {P.}~\bibnamefont {Simonetti}},\ }\href
  {\doibase 10.1103/PhysRevB.57.4694} {\bibfield  {journal} {\bibinfo
  {journal} {Phys. Rev. B}\ }\textbf {\bibinfo {volume} {57}},\ \bibinfo
  {pages} {4694} (\bibinfo {year} {1998})}\BibitemShut {NoStop}%
\bibitem [{\citenamefont {Ameduri}\ \emph {et~al.}(1999)\citenamefont
  {Ameduri}, \citenamefont {Efthimiou},\ and\ \citenamefont
  {Gerganov}}]{Ameduri1998a}%
  \BibitemOpen
  \bibfield  {author} {\bibinfo {author} {\bibfnamefont {M.}~\bibnamefont
  {Ameduri}}, \bibinfo {author} {\bibfnamefont {C.~J.}\ \bibnamefont
  {Efthimiou}}, \ and\ \bibinfo {author} {\bibfnamefont {B.}~\bibnamefont
  {Gerganov}},\ }\href {\doibase 10.1142/S021773239900242X} {\bibfield
  {journal} {\bibinfo  {journal} {Mod. Phys. Lett.}\ }\textbf {\bibinfo
  {volume} {A14}},\ \bibinfo {pages} {2341} (\bibinfo {year} {1999})},\ \Eprint
  {http://arxiv.org/abs/hep-th/9810184} {arXiv:hep-th/9810184 [hep-th]}
  \BibitemShut {NoStop}%
\bibitem [{\citenamefont {Bazhanov}\ \emph {et~al.}(2018)\citenamefont
  {Bazhanov}, \citenamefont {Lukyanov},\ and\ \citenamefont
  {Runov}}]{Bazhanov2018}%
  \BibitemOpen
  \bibfield  {author} {\bibinfo {author} {\bibfnamefont {V.~V.}\ \bibnamefont
  {Bazhanov}}, \bibinfo {author} {\bibfnamefont {S.~L.}\ \bibnamefont
  {Lukyanov}}, \ and\ \bibinfo {author} {\bibfnamefont {B.~A.}\ \bibnamefont
  {Runov}},\ }\href {\doibase https://doi.org/10.1016/j.nuclphysb.2017.12.022}
  {\bibfield  {journal} {\bibinfo  {journal} {Nuclear Physics B}\ }\textbf
  {\bibinfo {volume} {927}},\ \bibinfo {pages} {468 } (\bibinfo {year}
  {2018})}\BibitemShut {NoStop}%
\bibitem [{\citenamefont {Fateev}\ \emph {et~al.}(1993)\citenamefont {Fateev},
  \citenamefont {Onofri},\ and\ \citenamefont {Zamolodchikov}}]{Fateev1993}%
  \BibitemOpen
  \bibfield  {author} {\bibinfo {author} {\bibfnamefont {V.}~\bibnamefont
  {Fateev}}, \bibinfo {author} {\bibfnamefont {E.}~\bibnamefont {Onofri}}, \
  and\ \bibinfo {author} {\bibfnamefont {A.}~\bibnamefont {Zamolodchikov}},\
  }\href {\doibase https://doi.org/10.1016/0550-3213(93)90001-6} {\bibfield
  {journal} {\bibinfo  {journal} {Nuclear Physics B}\ }\textbf {\bibinfo
  {volume} {406}},\ \bibinfo {pages} {521 } (\bibinfo {year}
  {1993})}\BibitemShut {NoStop}%
\bibitem [{sol()}]{sol_foot}%
  \BibitemOpen
  \href@noop {} {}\bibinfo {note} {On the integrable manifold, the mass of the
  $\varphi_2$ soliton is also given by $m_s$.}\BibitemShut {Stop}%
\bibitem [{\citenamefont {Klassen}\ and\ \citenamefont
  {Melzer}(1991)}]{Klassen1991}%
  \BibitemOpen
  \bibfield  {author} {\bibinfo {author} {\bibfnamefont {T.~R.}\ \bibnamefont
  {Klassen}}\ and\ \bibinfo {author} {\bibfnamefont {E.}~\bibnamefont
  {Melzer}},\ }\href {\doibase https://doi.org/10.1016/0550-3213(91)90159-U}
  {\bibfield  {journal} {\bibinfo  {journal} {Nuclear Physics B}\ }\textbf
  {\bibinfo {volume} {350}},\ \bibinfo {pages} {635 } (\bibinfo {year}
  {1991})}\BibitemShut {NoStop}%
\bibitem [{\citenamefont {Dashen}\ \emph {et~al.}(1969)\citenamefont {Dashen},
  \citenamefont {Ma},\ and\ \citenamefont {Bernstein}}]{Dashen1969}%
  \BibitemOpen
  \bibfield  {author} {\bibinfo {author} {\bibfnamefont {R.}~\bibnamefont
  {Dashen}}, \bibinfo {author} {\bibfnamefont {S.-k.}\ \bibnamefont {Ma}}, \
  and\ \bibinfo {author} {\bibfnamefont {H.~J.}\ \bibnamefont {Bernstein}},\
  }\href {\doibase 10.1103/PhysRev.187.345} {\bibfield  {journal} {\bibinfo
  {journal} {Phys. Rev.}\ }\textbf {\bibinfo {volume} {187}},\ \bibinfo {pages}
  {345} (\bibinfo {year} {1969})}\BibitemShut {NoStop}%
\bibitem [{\citenamefont {Takahashi}\ and\ \citenamefont
  {Suzuki}(1972)}]{Takahashi1972}%
  \BibitemOpen
  \bibfield  {author} {\bibinfo {author} {\bibfnamefont {M.}~\bibnamefont
  {Takahashi}}\ and\ \bibinfo {author} {\bibfnamefont {M.}~\bibnamefont
  {Suzuki}},\ }\href {\doibase 10.1143/PTP.48.2187} {\bibfield  {journal}
  {\bibinfo  {journal} {Progress of Theoretical Physics}\ }\textbf {\bibinfo
  {volume} {48}},\ \bibinfo {pages} {2187} (\bibinfo {year} {1972})},\ \Eprint
  {http://arxiv.org/abs/http://oup.prod.sis.lan/ptp/article-pdf/48/6/2187/5255323/48-6-2187.pdf}
  {http://oup.prod.sis.lan/ptp/article-pdf/48/6/2187/5255323/48-6-2187.pdf}
  \BibitemShut {NoStop}%
\bibitem [{\citenamefont {Fendley}\ and\ \citenamefont
  {Intriligator}(1992)}]{Fendley1992}%
  \BibitemOpen
  \bibfield  {author} {\bibinfo {author} {\bibfnamefont {P.}~\bibnamefont
  {Fendley}}\ and\ \bibinfo {author} {\bibfnamefont {K.}~\bibnamefont
  {Intriligator}},\ }\href {\doibase
  https://doi.org/10.1016/0550-3213(92)90365-I} {\bibfield  {journal} {\bibinfo
   {journal} {Nuclear Physics B}\ }\textbf {\bibinfo {volume} {372}},\ \bibinfo
  {pages} {533 } (\bibinfo {year} {1992})}\BibitemShut {NoStop}%
\bibitem [{\citenamefont {Takahashi}(2005)}]{Takahashi2005}%
  \BibitemOpen
  \bibfield  {author} {\bibinfo {author} {\bibfnamefont {M.}~\bibnamefont
  {Takahashi}},\ }\href {https://books.google.de/books?id=ubGcM-JCT0IC} {\emph
  {\bibinfo {title} {Thermodynamics of One-Dimensional Solvable Models}}}\
  (\bibinfo  {publisher} {Cambridge University Press},\ \bibinfo {year}
  {2005})\BibitemShut {NoStop}%
\bibitem [{\citenamefont {Korepin}\ \emph {et~al.}(1993)\citenamefont
  {Korepin}, \citenamefont {Izergin},\ and\ \citenamefont
  {Bogoliubov}}]{Korepin1993}%
  \BibitemOpen
  \bibfield  {author} {\bibinfo {author} {\bibfnamefont {V.~E.}\ \bibnamefont
  {Korepin}}, \bibinfo {author} {\bibfnamefont {a.~G.}\ \bibnamefont
  {Izergin}}, \ and\ \bibinfo {author} {\bibfnamefont {N.~M.}\ \bibnamefont
  {Bogoliubov}},\ }\href {http://arxiv.org/abs/cond-mat/9301031} {\enquote
  {\bibinfo {title} {{Quantum Inverse Scattering Method and Correlation
  Functions}},}\ } (\bibinfo {year} {1993})\BibitemShut {NoStop}%
\bibitem [{\citenamefont {Faddeev}(1996)}]{Faddeev1996}%
  \BibitemOpen
  \bibfield  {author} {\bibinfo {author} {\bibfnamefont {L.~D.}\ \bibnamefont
  {Faddeev}},\ }in\ \href@noop {} {\emph {\bibinfo {booktitle} {{Relativistic
  gravitation and gravitational radiation. Proceedings, School of Physics, Les
  Houches, France, September 26-October 6, 1995}}}}\ (\bibinfo {year} {1996})\
  pp.\ \bibinfo {pages} {pp. 149--219},\ \Eprint
  {http://arxiv.org/abs/hep-th/9605187} {arXiv:hep-th/9605187 [hep-th]}
  \BibitemShut {NoStop}%
\bibitem [{\citenamefont
  {Zamolodchikov}(1991{\natexlab{a}})}]{Zamolodchikov1991a}%
  \BibitemOpen
  \bibfield  {author} {\bibinfo {author} {\bibfnamefont {A.}~\bibnamefont
  {Zamolodchikov}},\ }\href {\doibase
  https://doi.org/10.1016/0550-3213(91)90422-T} {\bibfield  {journal} {\bibinfo
   {journal} {Nuclear Physics B}\ }\textbf {\bibinfo {volume} {358}},\ \bibinfo
  {pages} {497 } (\bibinfo {year} {1991}{\natexlab{a}})}\BibitemShut {NoStop}%
\bibitem [{\citenamefont {Fendley}\ \emph {et~al.}(1993)\citenamefont
  {Fendley}, \citenamefont {Saleur},\ and\ \citenamefont
  {Zamolodchikov}}]{Fendley1993}%
  \BibitemOpen
  \bibfield  {author} {\bibinfo {author} {\bibfnamefont {P.}~\bibnamefont
  {Fendley}}, \bibinfo {author} {\bibfnamefont {H.}~\bibnamefont {Saleur}}, \
  and\ \bibinfo {author} {\bibfnamefont {A.~B.}\ \bibnamefont
  {Zamolodchikov}},\ }\href {\doibase 10.1142/S0217751X93002265} {\bibfield
  {journal} {\bibinfo  {journal} {International Journal of Modern Physics A}\
  }\textbf {\bibinfo {volume} {08}},\ \bibinfo {pages} {5717} (\bibinfo {year}
  {1993})},\ \Eprint
  {http://arxiv.org/abs/https://doi.org/10.1142/S0217751X93002265}
  {https://doi.org/10.1142/S0217751X93002265} \BibitemShut {NoStop}%
\bibitem [{\citenamefont {Zamolodchikov}(2006)}]{Zamolodchikov2006}%
  \BibitemOpen
  \bibfield  {author} {\bibinfo {author} {\bibfnamefont {A.~B.}\ \bibnamefont
  {Zamolodchikov}},\ }\href {\doibase 10.1088/0305-4470/39/41/s08} {\bibfield
  {journal} {\bibinfo  {journal} {Journal of Physics A: Mathematical and
  General}\ }\textbf {\bibinfo {volume} {39}},\ \bibinfo {pages} {12847}
  (\bibinfo {year} {2006})}\BibitemShut {NoStop}%
\bibitem [{\citenamefont {Manucharyan}\ \emph {et~al.}(2009)\citenamefont
  {Manucharyan}, \citenamefont {Koch}, \citenamefont {Glazman},\ and\
  \citenamefont {Devoret}}]{Manucharyan2009}%
  \BibitemOpen
  \bibfield  {author} {\bibinfo {author} {\bibfnamefont {V.~E.}\ \bibnamefont
  {Manucharyan}}, \bibinfo {author} {\bibfnamefont {J.}~\bibnamefont {Koch}},
  \bibinfo {author} {\bibfnamefont {L.~I.}\ \bibnamefont {Glazman}}, \ and\
  \bibinfo {author} {\bibfnamefont {M.~H.}\ \bibnamefont {Devoret}},\ }\href
  {\doibase 10.1126/science.1175552} {\bibfield  {journal} {\bibinfo  {journal}
  {Science}\ }\textbf {\bibinfo {volume} {326}},\ \bibinfo {pages} {113}
  (\bibinfo {year} {2009})},\ \Eprint
  {http://arxiv.org/abs/http://science.sciencemag.org/content/326/5949/113.full.pdf}
  {http://science.sciencemag.org/content/326/5949/113.full.pdf} \BibitemShut
  {NoStop}%
\bibitem [{\citenamefont {Koch}\ \emph {et~al.}(2009)\citenamefont {Koch},
  \citenamefont {Manucharyan}, \citenamefont {Devoret},\ and\ \citenamefont
  {Glazman}}]{Koch2009}%
  \BibitemOpen
  \bibfield  {author} {\bibinfo {author} {\bibfnamefont {J.}~\bibnamefont
  {Koch}}, \bibinfo {author} {\bibfnamefont {V.}~\bibnamefont {Manucharyan}},
  \bibinfo {author} {\bibfnamefont {M.~H.}\ \bibnamefont {Devoret}}, \ and\
  \bibinfo {author} {\bibfnamefont {L.~I.}\ \bibnamefont {Glazman}},\ }\href
  {\doibase 10.1103/PhysRevLett.103.217004} {\bibfield  {journal} {\bibinfo
  {journal} {Phys. Rev. Lett.}\ }\textbf {\bibinfo {volume} {103}},\ \bibinfo
  {pages} {217004} (\bibinfo {year} {2009})}\BibitemShut {NoStop}%
\bibitem [{\citenamefont {Le~Hur}(2012)}]{LeHur2012}%
  \BibitemOpen
  \bibfield  {author} {\bibinfo {author} {\bibfnamefont {K.}~\bibnamefont
  {Le~Hur}},\ }\href {\doibase 10.1103/PhysRevB.85.140506} {\bibfield
  {journal} {\bibinfo  {journal} {Phys. Rev. B}\ }\textbf {\bibinfo {volume}
  {85}},\ \bibinfo {pages} {140506} (\bibinfo {year} {2012})}\BibitemShut
  {NoStop}%
\bibitem [{\citenamefont {Goldstein}\ \emph {et~al.}(2013)\citenamefont
  {Goldstein}, \citenamefont {Devoret}, \citenamefont {Houzet},\ and\
  \citenamefont {Glazman}}]{Goldstein2013}%
  \BibitemOpen
  \bibfield  {author} {\bibinfo {author} {\bibfnamefont {M.}~\bibnamefont
  {Goldstein}}, \bibinfo {author} {\bibfnamefont {M.~H.}\ \bibnamefont
  {Devoret}}, \bibinfo {author} {\bibfnamefont {M.}~\bibnamefont {Houzet}}, \
  and\ \bibinfo {author} {\bibfnamefont {L.~I.}\ \bibnamefont {Glazman}},\
  }\href {\doibase 10.1103/PhysRevLett.110.017002} {\bibfield  {journal}
  {\bibinfo  {journal} {Phys. Rev. Lett.}\ }\textbf {\bibinfo {volume} {110}},\
  \bibinfo {pages} {017002} (\bibinfo {year} {2013})}\BibitemShut {NoStop}%
\bibitem [{\citenamefont {Koch}\ \emph {et~al.}(2007)\citenamefont {Koch},
  \citenamefont {Yu}, \citenamefont {Gambetta}, \citenamefont {Houck},
  \citenamefont {Schuster}, \citenamefont {Majer}, \citenamefont {Blais},
  \citenamefont {Devoret}, \citenamefont {Girvin},\ and\ \citenamefont
  {Schoelkopf}}]{Koch_Schoelkopf_2007}%
  \BibitemOpen
  \bibfield  {author} {\bibinfo {author} {\bibfnamefont {J.}~\bibnamefont
  {Koch}}, \bibinfo {author} {\bibfnamefont {T.~M.}\ \bibnamefont {Yu}},
  \bibinfo {author} {\bibfnamefont {J.}~\bibnamefont {Gambetta}}, \bibinfo
  {author} {\bibfnamefont {A.~A.}\ \bibnamefont {Houck}}, \bibinfo {author}
  {\bibfnamefont {D.~I.}\ \bibnamefont {Schuster}}, \bibinfo {author}
  {\bibfnamefont {J.}~\bibnamefont {Majer}}, \bibinfo {author} {\bibfnamefont
  {A.}~\bibnamefont {Blais}}, \bibinfo {author} {\bibfnamefont {M.~H.}\
  \bibnamefont {Devoret}}, \bibinfo {author} {\bibfnamefont {S.~M.}\
  \bibnamefont {Girvin}}, \ and\ \bibinfo {author} {\bibfnamefont {R.~J.}\
  \bibnamefont {Schoelkopf}},\ }\href {\doibase 10.1103/PhysRevA.76.042319}
  {\bibfield  {journal} {\bibinfo  {journal} {Phys. Rev. A}\ }\textbf {\bibinfo
  {volume} {76}},\ \bibinfo {pages} {1} (\bibinfo {year} {2007})}\BibitemShut
  {NoStop}%
\bibitem [{\citenamefont {Guarcello}\ \emph
  {et~al.}(2018{\natexlab{a}})\citenamefont {Guarcello}, \citenamefont
  {Solinas}, \citenamefont {Braggio},\ and\ \citenamefont
  {Giazotto}}]{Guarcello2018}%
  \BibitemOpen
  \bibfield  {author} {\bibinfo {author} {\bibfnamefont {C.}~\bibnamefont
  {Guarcello}}, \bibinfo {author} {\bibfnamefont {P.}~\bibnamefont {Solinas}},
  \bibinfo {author} {\bibfnamefont {A.}~\bibnamefont {Braggio}}, \ and\
  \bibinfo {author} {\bibfnamefont {F.}~\bibnamefont {Giazotto}},\ }\href
  {\doibase 10.1038/s41598-018-30268-1} {\bibfield  {journal} {\bibinfo
  {journal} {Scientific Reports}\ }\textbf {\bibinfo {volume} {8}},\ \bibinfo
  {pages} {12287} (\bibinfo {year} {2018}{\natexlab{a}})}\BibitemShut {NoStop}%
\bibitem [{\citenamefont {Bergeal}\ \emph {et~al.}(2010)\citenamefont
  {Bergeal}, \citenamefont {Vijay}, \citenamefont {Manucharyan}, \citenamefont
  {Siddiqi}, \citenamefont {Schoelkopf}, \citenamefont {Girvin},\ and\
  \citenamefont {Devoret}}]{Bergeal_Devoret_2010}%
  \BibitemOpen
  \bibfield  {author} {\bibinfo {author} {\bibfnamefont {N.}~\bibnamefont
  {Bergeal}}, \bibinfo {author} {\bibfnamefont {R.}~\bibnamefont {Vijay}},
  \bibinfo {author} {\bibfnamefont {V.~E.}\ \bibnamefont {Manucharyan}},
  \bibinfo {author} {\bibfnamefont {I.}~\bibnamefont {Siddiqi}}, \bibinfo
  {author} {\bibfnamefont {R.~J.}\ \bibnamefont {Schoelkopf}}, \bibinfo
  {author} {\bibfnamefont {S.~M.}\ \bibnamefont {Girvin}}, \ and\ \bibinfo
  {author} {\bibfnamefont {M.~H.}\ \bibnamefont {Devoret}},\ }\href
  {http://dx.doi.org/10.1038/nphys1516} {\bibfield  {journal} {\bibinfo
  {journal} {Nat Phys}\ }\textbf {\bibinfo {volume} {6}},\ \bibinfo {pages}
  {296} (\bibinfo {year} {2010})}\BibitemShut {NoStop}%
\bibitem [{\citenamefont {Roy}\ \emph {et~al.}(2015)\citenamefont {Roy},
  \citenamefont {Jiang}, \citenamefont {Stone},\ and\ \citenamefont
  {Devoret}}]{Roy_Devoret_2015}%
  \BibitemOpen
  \bibfield  {author} {\bibinfo {author} {\bibfnamefont {A.}~\bibnamefont
  {Roy}}, \bibinfo {author} {\bibfnamefont {L.}~\bibnamefont {Jiang}}, \bibinfo
  {author} {\bibfnamefont {A.~D.}\ \bibnamefont {Stone}}, \ and\ \bibinfo
  {author} {\bibfnamefont {M.}~\bibnamefont {Devoret}},\ }\href {\doibase
  10.1103/PhysRevLett.115.150503} {\bibfield  {journal} {\bibinfo  {journal}
  {Phys. Rev. Lett.}\ }\textbf {\bibinfo {volume} {115}},\ \bibinfo {pages}
  {150503} (\bibinfo {year} {2015})}\BibitemShut {NoStop}%
\bibitem [{fou()}]{four_mode_foot}%
  \BibitemOpen
  \href@noop {} {}\bibinfo {note} {A four-field integrable QFT with QEC-s will
  be described elsewhere.}\BibitemShut {Stop}%
\bibitem [{\citenamefont {Fu}(2010)}]{Fu2010}%
  \BibitemOpen
  \bibfield  {author} {\bibinfo {author} {\bibfnamefont {L.}~\bibnamefont
  {Fu}},\ }\href {\doibase 10.1103/PhysRevLett.104.056402} {\bibfield
  {journal} {\bibinfo  {journal} {Phys. Rev. Lett.}\ }\textbf {\bibinfo
  {volume} {104}},\ \bibinfo {pages} {1} (\bibinfo {year} {2010})}\BibitemShut
  {NoStop}%
\bibitem [{\citenamefont {Zamolodchikov}(1981)}]{Zamolodchikov1981}%
  \BibitemOpen
  \bibfield  {author} {\bibinfo {author} {\bibfnamefont {A.~B.}\ \bibnamefont
  {Zamolodchikov}},\ }\href
  {https://projecteuclid.org:443/euclid.cmp/1103909139} {\bibfield  {journal}
  {\bibinfo  {journal} {Comm. Math. Phys.}\ }\textbf {\bibinfo {volume} {79}},\
  \bibinfo {pages} {489} (\bibinfo {year} {1981})}\BibitemShut {NoStop}%
\bibitem [{\citenamefont {Zamolodchikov}(1995)}]{Zamolodchikov1995}%
  \BibitemOpen
  \bibfield  {author} {\bibinfo {author} {\bibfnamefont {A.~B.}\ \bibnamefont
  {Zamolodchikov}},\ }\href {\doibase 10.1142/S0217751X9500053X} {\bibfield
  {journal} {\bibinfo  {journal} {International Journal of Modern Physics A}\
  }\textbf {\bibinfo {volume} {10}},\ \bibinfo {pages} {1125} (\bibinfo {year}
  {1995})},\ \Eprint
  {http://arxiv.org/abs/https://doi.org/10.1142/S0217751X9500053X}
  {https://doi.org/10.1142/S0217751X9500053X} \BibitemShut {NoStop}%
\bibitem [{\citenamefont {Polyakov}\ and\ \citenamefont
  {Wiegmann}(1983)}]{Polyakov1983}%
  \BibitemOpen
  \bibfield  {author} {\bibinfo {author} {\bibfnamefont {A.}~\bibnamefont
  {Polyakov}}\ and\ \bibinfo {author} {\bibfnamefont {P.}~\bibnamefont
  {Wiegmann}},\ }\href {\doibase https://doi.org/10.1016/0370-2693(83)91104-8}
  {\bibfield  {journal} {\bibinfo  {journal} {Physics Letters B}\ }\textbf
  {\bibinfo {volume} {131}},\ \bibinfo {pages} {121 } (\bibinfo {year}
  {1983})}\BibitemShut {NoStop}%
\bibitem [{\citenamefont {{Wiegmann}}(1985)}]{Wiegmann1984}%
  \BibitemOpen
  \bibfield  {author} {\bibinfo {author} {\bibfnamefont {P.~B.}\ \bibnamefont
  {{Wiegmann}}},\ }\href {\doibase 10.1016/0370-2693(85)91171-2} {\bibfield
  {journal} {\bibinfo  {journal} {Physics Letters B}\ }\textbf {\bibinfo
  {volume} {152}},\ \bibinfo {pages} {209} (\bibinfo {year}
  {1985})}\BibitemShut {NoStop}%
\bibitem [{\citenamefont
  {Zamolodchikov}(1991{\natexlab{b}})}]{Zamolodchikov1991b}%
  \BibitemOpen
  \bibfield  {author} {\bibinfo {author} {\bibfnamefont {A.}~\bibnamefont
  {Zamolodchikov}},\ }\href {\doibase
  https://doi.org/10.1016/0550-3213(91)90423-U} {\bibfield  {journal} {\bibinfo
   {journal} {Nuclear Physics B}\ }\textbf {\bibinfo {volume} {358}},\ \bibinfo
  {pages} {524 } (\bibinfo {year} {1991}{\natexlab{b}})}\BibitemShut {NoStop}%
\bibitem [{\citenamefont {Lewin}(1981)}]{Lewin1981}%
  \BibitemOpen
  \bibfield  {author} {\bibinfo {author} {\bibfnamefont {L.}~\bibnamefont
  {Lewin}},\ }\href {https://books.google.de/books?id=yETvAAAAMAAJ} {\emph
  {\bibinfo {title} {Polylogarithms and associated functions}}}\ (\bibinfo
  {publisher} {North Holland},\ \bibinfo {year} {1981})\BibitemShut {NoStop}%
\bibitem [{\citenamefont {Clerk}\ \emph {et~al.}(2010)\citenamefont {Clerk},
  \citenamefont {Devoret}, \citenamefont {Girvin}, \citenamefont {Marquardt},\
  and\ \citenamefont {Schoelkopf}}]{Clerk_Schoelkopf_2010}%
  \BibitemOpen
  \bibfield  {author} {\bibinfo {author} {\bibfnamefont {A.~A.}\ \bibnamefont
  {Clerk}}, \bibinfo {author} {\bibfnamefont {M.~H.}\ \bibnamefont {Devoret}},
  \bibinfo {author} {\bibfnamefont {S.~M.}\ \bibnamefont {Girvin}}, \bibinfo
  {author} {\bibfnamefont {F.}~\bibnamefont {Marquardt}}, \ and\ \bibinfo
  {author} {\bibfnamefont {R.~J.}\ \bibnamefont {Schoelkopf}},\ }\href
  {\doibase 10.1103/RevModPhys.82.1155} {\bibfield  {journal} {\bibinfo
  {journal} {Rev. Mod. Phys.}\ }\textbf {\bibinfo {volume} {82}},\ \bibinfo
  {pages} {1155} (\bibinfo {year} {2010})}\BibitemShut {NoStop}%
\bibitem [{\citenamefont {Roy}\ and\ \citenamefont {Devoret}(2016)}]{Roy2016}%
  \BibitemOpen
  \bibfield  {author} {\bibinfo {author} {\bibfnamefont {A.}~\bibnamefont
  {Roy}}\ and\ \bibinfo {author} {\bibfnamefont {M.}~\bibnamefont {Devoret}},\
  }\href {\doibase https://doi.org/10.1016/j.crhy.2016.07.012} {\bibfield
  {journal} {\bibinfo  {journal} {Comptes Rendus Physique}\ }\textbf {\bibinfo
  {volume} {17}},\ \bibinfo {pages} {740 } (\bibinfo {year}
  {2016})}\BibitemShut {NoStop}%
\bibitem [{\citenamefont {Roy}\ and\ \citenamefont {Devoret}(2018)}]{Roy2018a}%
  \BibitemOpen
  \bibfield  {author} {\bibinfo {author} {\bibfnamefont {A.}~\bibnamefont
  {Roy}}\ and\ \bibinfo {author} {\bibfnamefont {M.}~\bibnamefont {Devoret}},\
  }\href {\doibase 10.1103/PhysRevB.98.045405} {\bibfield  {journal} {\bibinfo
  {journal} {Phys. Rev. B}\ }\textbf {\bibinfo {volume} {98}},\ \bibinfo
  {pages} {045405} (\bibinfo {year} {2018})}\BibitemShut {NoStop}%
\bibitem [{\citenamefont {{Kuzmin}}\ \emph {et~al.}(2018)\citenamefont
  {{Kuzmin}}, \citenamefont {{Grabon}}, \citenamefont {{Mehta}}, \citenamefont
  {{Mencia}}, \citenamefont {{Pankratova}}, \citenamefont {{Goldstein}},\ and\
  \citenamefont {{Manucharyan}}}]{Kuzmin2018}%
  \BibitemOpen
  \bibfield  {author} {\bibinfo {author} {\bibfnamefont {R.}~\bibnamefont
  {{Kuzmin}}}, \bibinfo {author} {\bibfnamefont {N.}~\bibnamefont {{Grabon}}},
  \bibinfo {author} {\bibfnamefont {N.}~\bibnamefont {{Mehta}}}, \bibinfo
  {author} {\bibfnamefont {R.}~\bibnamefont {{Mencia}}}, \bibinfo {author}
  {\bibfnamefont {N.}~\bibnamefont {{Pankratova}}}, \bibinfo {author}
  {\bibfnamefont {M.}~\bibnamefont {{Goldstein}}}, \ and\ \bibinfo {author}
  {\bibfnamefont {V.}~\bibnamefont {{Manucharyan}}},\ }in\ \href@noop {} {\emph
  {\bibinfo {booktitle} {APS Meeting Abstracts}}}\ (\bibinfo {year} {2018})\
  p.\ \bibinfo {pages} {H33.003}\BibitemShut {NoStop}%
\bibitem [{\citenamefont {Fominaya}\ \emph {et~al.}(1997)\citenamefont
  {Fominaya}, \citenamefont {Fournier}, \citenamefont {Gandit},\ and\
  \citenamefont {Chaussy}}]{Fominaya1997}%
  \BibitemOpen
  \bibfield  {author} {\bibinfo {author} {\bibfnamefont {F.}~\bibnamefont
  {Fominaya}}, \bibinfo {author} {\bibfnamefont {T.}~\bibnamefont {Fournier}},
  \bibinfo {author} {\bibfnamefont {P.}~\bibnamefont {Gandit}}, \ and\ \bibinfo
  {author} {\bibfnamefont {J.}~\bibnamefont {Chaussy}},\ }\href {\doibase
  10.1063/1.1148366} {\bibfield  {journal} {\bibinfo  {journal} {Review of
  Scientific Instruments}\ }\textbf {\bibinfo {volume} {68}},\ \bibinfo {pages}
  {4191} (\bibinfo {year} {1997})},\ \Eprint
  {http://arxiv.org/abs/https://doi.org/10.1063/1.1148366}
  {https://doi.org/10.1063/1.1148366} \BibitemShut {NoStop}%
\bibitem [{\citenamefont {Rabani}\ \emph {et~al.}(2008)\citenamefont {Rabani},
  \citenamefont {Taddei}, \citenamefont {Bourgeois}, \citenamefont {Fazio},\
  and\ \citenamefont {Giazotto}}]{Rabani2008}%
  \BibitemOpen
  \bibfield  {author} {\bibinfo {author} {\bibfnamefont {H.}~\bibnamefont
  {Rabani}}, \bibinfo {author} {\bibfnamefont {F.}~\bibnamefont {Taddei}},
  \bibinfo {author} {\bibfnamefont {O.}~\bibnamefont {Bourgeois}}, \bibinfo
  {author} {\bibfnamefont {R.}~\bibnamefont {Fazio}}, \ and\ \bibinfo {author}
  {\bibfnamefont {F.}~\bibnamefont {Giazotto}},\ }\href {\doibase
  10.1103/PhysRevB.78.012503} {\bibfield  {journal} {\bibinfo  {journal} {Phys.
  Rev. B}\ }\textbf {\bibinfo {volume} {78}},\ \bibinfo {pages} {012503}
  (\bibinfo {year} {2008})}\BibitemShut {NoStop}%
\bibitem [{\citenamefont {Guarcello}\ \emph
  {et~al.}(2018{\natexlab{b}})\citenamefont {Guarcello}, \citenamefont
  {Solinas}, \citenamefont {Braggio},\ and\ \citenamefont
  {Giazotto}}]{Guarcello2018a}%
  \BibitemOpen
  \bibfield  {author} {\bibinfo {author} {\bibfnamefont {C.}~\bibnamefont
  {Guarcello}}, \bibinfo {author} {\bibfnamefont {P.}~\bibnamefont {Solinas}},
  \bibinfo {author} {\bibfnamefont {A.}~\bibnamefont {Braggio}}, \ and\
  \bibinfo {author} {\bibfnamefont {F.}~\bibnamefont {Giazotto}},\ }\href
  {\doibase 10.1103/PhysRevB.98.104501} {\bibfield  {journal} {\bibinfo
  {journal} {Phys. Rev. B}\ }\textbf {\bibinfo {volume} {98}},\ \bibinfo
  {pages} {104501} (\bibinfo {year} {2018}{\natexlab{b}})}\BibitemShut
  {NoStop}%
\bibitem [{\citenamefont {Hida}\ and\ \citenamefont {Eckern}(1984)}]{Hida1984}%
  \BibitemOpen
  \bibfield  {author} {\bibinfo {author} {\bibfnamefont {K.}~\bibnamefont
  {Hida}}\ and\ \bibinfo {author} {\bibfnamefont {U.}~\bibnamefont {Eckern}},\
  }\href {\doibase 10.1103/PhysRevB.30.4096} {\bibfield  {journal} {\bibinfo
  {journal} {Phys. Rev. B}\ }\textbf {\bibinfo {volume} {30}},\ \bibinfo
  {pages} {4096} (\bibinfo {year} {1984})}\BibitemShut {NoStop}%
\bibitem [{\citenamefont {{Hida}}(1985)}]{Hida1985}%
  \BibitemOpen
  \bibfield  {author} {\bibinfo {author} {\bibfnamefont {K.}~\bibnamefont
  {{Hida}}},\ }\href {\doibase 10.1007/BF01307781} {\bibfield  {journal}
  {\bibinfo  {journal} {Zeitschrift fur Physik B Condensed Matter}\ }\textbf
  {\bibinfo {volume} {61}},\ \bibinfo {pages} {223} (\bibinfo {year}
  {1985})}\BibitemShut {NoStop}%
\end{thebibliography}%

\end{document}